# Interference effects in polarization-controlled Rayleigh scattering in twisted bilayer graphene

Disha Arora, Deepanshu Aggarwal, Sankalpa Ghosh and Rohit Narula
*Department of Physics, Indian Institute of Technology Delhi, New Delhi-110016, India*

We calculate the polarization-controlled Rayleigh scattering response of twisted bilayer graphene (tBLG) based on the continuum electronic band model developed by Bistritzer and MacDonald while considering its refinements which address the effects of structural corrugation, doping-dependent Hartree interactions and particle-hole asymmetry. The dominant wave vectors for the Rayleigh scattering process emanate from various regions of the Moiré Brillouin zone (MBZ) in contrast to single-layer graphene (SLG) and AB-stacked bilayer graphene (AB-BLG), where the dominant contributions always stem from the vicinity of the $\boldsymbol{K}$ point for optical laser energies and below. Compared to SLG, the integrated Rayleigh intensity is strongly enhanced for small twist angles (*e.g.*, at a twist angle $\theta = 1.2°$, the integrated Rayleigh intensity at laser energy $E_l = 2$ eV enhances by a factor of $\sim 100$ for the case of parallel polarization). While for the case of cross-polarization, it exhibits a markedly complex behavior suggestive of strong interference effects mediated by the optical matrix elements. We find that at small twist angles, *e.g.*, $\theta = 1.05°$, the corrugation effects strongly enhances the ratio $\boldsymbol{R}_A = \frac{\text{integrated Rayleigh intensity for parallel polarization}}{\text{integrated Rayleigh intensity for cross-polarization}}$ by $\sim 1300$ times *viz a viz* SLG or AB-BLG. Measured as a function of the incoming laser energy $E_l$, $\boldsymbol{R}_A$ exhibits a characteristic evolution as the twist angle reduces, thus providing a unique fingerprint of the prevailing twist angle of the tBLG sample under study, which would be interesting to verify experimentally.

## I. INTRODUCTION

While the two-dimensional superlattice: twisted bilayer graphene (tBLG) was fabricated nearly a decade ago [1–5], it has again garnered attention for its ability to host strongly correlated phases, particularly at the so-called magic angles of twist or misorientation $\theta$. These include the observations of alternating superconducting [6–8] and interaction-induced insulating phases [9–15], magnetism [16, 17], linear-in-temperature low-temperature resistivity and anomalous quantum hall state [18–20] etc. Experimental studies also reveal that the twisting between the two layers can greatly enhance the optoelectronic properties like optical conductivity [21–24], photo luminescence [25, 26], optical absorption [27], and photocurrent [28]. In the literature, there are various theoretical models [29–38] demonstrating that as $\theta$ is reduced, the interference of two lattice periods results in the formation of a Moiré pattern with a long wavelength, where features like band gaps and Van Hove singularities appear in the far-infrared spectrum, and the band velocity of the Dirac cone is significantly decreased [39, 40].

Spectroscopic techniques such as Rayleigh (elastic) [41–44] and Raman (inelastic) [45, 46] scattering provide an invaluable insight into the electronic [39, 47–49] and vibrational structure [50–52] of $sp^2$ type carbons [53–55]. Experimentally, the most prominent among them is Raman scattering which serves as a non-destructive technique for the ready identification of the structural properties in graphene and nearly all its conceivable aggregations [51], such as the number of layers, lattice-orientation [50], level of doping [56] and disorder [57, 58]. First-order Raman and double resonance Raman scattering have been extensively studied in $sp^2$ carbons [58–65]. The most prominent features observed in the Raman spectra of monolayer graphene are the $G$ and $2D$ bands appearing at 1582 cm$^{-1}$ and 2700 cm$^{-1}$, respectively, at a laser energy of 2.41 eV. A disorder-induced $D$-band is observed for a disordered sample at about half of the frequency of the $2D$ band [60]. The analysis of double-resonant Raman scattering in graphite shows that in the full integration of the Raman cross-section, the contributions by phonons from exactly the $\boldsymbol{K}$ point cancel due to destructive interference [66]. Indeed tBLG is no exception, and each twist angle is characterized by a unique set of phonon frequencies, which together provide a Raman signature, while their line widths provide a straightforward test for structural homogeneity [67]. In tBLG, the phonons are activated in the interior of the MBZ due to a $\theta$-dependent wave vector generated by the superlattice that is used to probe the phonon dispersion in tBLG, and despite the absence of a stacking arrangement in tBLG, layer breathing vibrations (namely the $ZO'$ phonons) are observed [68]. In a small range of twist angles where the intensity of the $G$ Raman peak is strongly enhanced, two new Raman modes (below 100 cm$^{-1}$) are observed. This suggests that these low energy modes and the $G$ Raman mode share the same resonance enhancement mechanism as a function of twist angle [69]. Recent research demonstrates that the linewidth of the $G$ band close to the magic angle is affected by the electron-phonon interaction regardless of laser excitation wavelength [70]. Nevertheless, the Raman-scattered photons are in the minority compared to those scattered elastically (Rayleigh) [71]. It remains to be established where the actual source of dominant contribution to the Rayleigh response of tBLG lies given its markedly complex manifold of electronic bands [29, 31, 32] as the probing laser energy $E_l$, and polarizer-analyzer orientation is varied. From a theoretical standpoint, it is also judicious to view the Rayleigh scattering phenomenon as elementary to all the higher-order processes involving light-matter interaction, which, besides the aforementioned Raman scattering, include, *e.g.*, one, two, three-photon scattering [72, 73].

In this paper, we delve into the Rayleigh or elastic scattering process in tBLG for a range of twist angles $\theta$, probing laser energies $E_l$, and polarizer-analyzer orientation combinations. For reference, we present a comparison with the Rayleigh scattering response of single-layer [75, 76] and AB-stacked bilayer graphene [77]. Our calculations show that in remarkable con-trast to SLG or AB-BLG, the dominant contribution to the Rayleigh scattering process in tBLG is not restricted to the wave vectors near the $\boldsymbol{K}$ point in the Brillouin zone. Instead, it highly depends on the twist angle and incoming laser energy and emanates from different parts of the Moiré Brillouin zone (MBZ). We also observe that for each twist angle, $\boldsymbol{R}_A$ exhibits a characteristic evolution with laser energy, which gives a unique fingerprint for experimental identification of different twist angles.

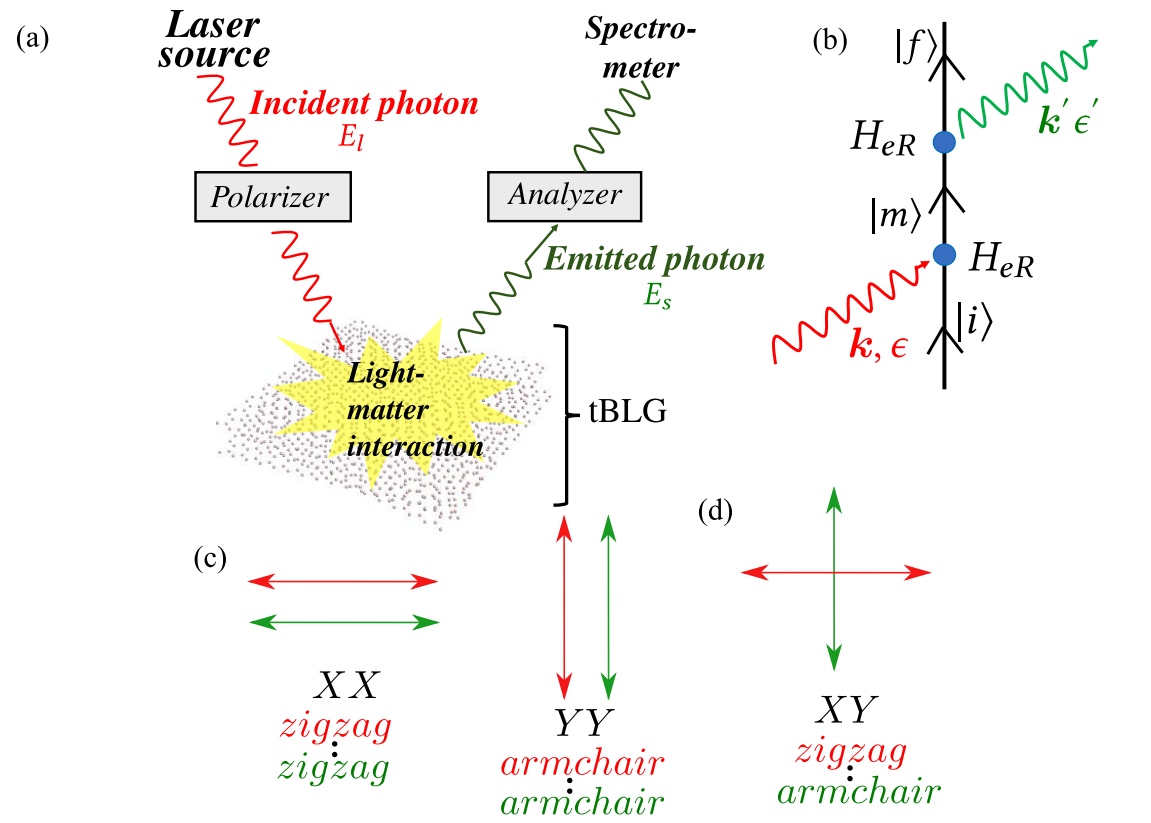

FIG. 1: (a) Schematic of one- photon absorption and emission (b) The corresponding Feynman diagram for Rayleigh scattering process. Red (green) arrows indicate the polarization directions of incoming (outgoing) light for (c) parallel ($XX$ or $YY$) and (b) cross-polarization ($XY$).

This paper is organized as follows: Sec. II discusses the theoretical background and provides the relevant expressions used in our calculations. The interaction of photons with tBLG is discussed and derived in Sec. III. Sec. IV discusses the behavior of optical matrix elements and provides a detailed analysis of the Rayleigh scattered signal for parallel and cross-polarization along with its dependency on the laser energy and twist angle. Our conclusions and scope of future work are outlined in Sec. V.

## II. THEORETICAL APPROACH

This section briefly reviews the theory of one-photon absorption and emission corresponding to Rayleigh scattering, which derives from second-order perturbation theory [73, 78]. Fig. 1 (a) shows the schematic of an experimental setup for the one-photon absorption and emission, and the Feynman diagram corresponding to Rayleigh scattering is shown in Fig. 1 (b). We have a laser source that emits photons of energy $E_l$, a polarizer and analyzer, the material system under consideration (tBLG in our case) and a spectrometer. The photons produced by the laser source pass through the polarizer and interact with the electronic subsystem of the material *via* the light-matter interaction Hamiltonian $H_{eR} = \boldsymbol{p}\cdot\boldsymbol{A}$. The electron makes a transition from its initial (ground) state, characterized by a state $|i\rangle$ to an excited state $|m\rangle$, eventually returning to its final state $|f\rangle$ by emitting photons of energy $E_s$. The emitted photons pass through the analyzer, which further constrains their optical polarization, and the output intensity is measured by the spectrometer. $k(k')$ and $\epsilon(\epsilon')$ represent the wave vector and polarization direction of the incident (emitted) photons. The red (green) arrow in Fig. 1 (c) and (d) represents the polarization direction of the incident (emitted) light for parallel ($XX$ or $YY$) and cross-polarization ($XY$), respectively. In the former, the polarizer and analyzer are aligned either along the zigzag: $X$ or armchair edge: $Y$ of an unrotated graphene sheet (refer Fig. 2 (a) for the orientation), while in the latter, incoming and outgoing light have a 90° out of phase difference between their optical polarizations [79], *e.g.*, if one is aligned along the zigzag edge, then the other is along the armchair edge (See Fig. 1 (d)).

From second-order perturbation theory, the transition matrix [73, 80] for the above process is given as

$$\mathcal{T}_{fi} = \sum_m \frac{\langle \psi_f | H_{eR} | \psi_m \rangle \langle \psi_m | H_{eR} | \psi_i \rangle}{(E_i + E_l - (E_m + \iota\gamma))} \tag{1}$$

and the transition probability rate is given by

$$w_{fi} = \frac{2\pi}{\hbar} \int_{\boldsymbol{k}} \left|\mathcal{T}_{fi}\right|^2 \delta(E_i + E_l - (E_f - E_s)) d^2\boldsymbol{k} \tag{2}$$

The domain of integration over $\boldsymbol{k}$ includes the entire Brillouin zone (BZ) of the crystalline material system under study. For the case of tBLG, the integration domain is over the so-called MBZ [5, 81] as defined by Bistritzer and MacDonald [29]. $\psi_i$, $\psi_f$ and $\psi_m$ represent the initial, final and intermediate states with energies $E_i$, $E_f$ and $E_m$, respectively, with $\gamma$ as the broadening parameter, which is taken to be a fixed fraction $\approx \frac{1}{20}th$ of the laser energy for all the considered transitions. $H_{eR}$ is the light-matter interaction Hamiltonian and is given by $\boldsymbol{p}\cdot\boldsymbol{A}$ where $\boldsymbol{p} = \hbar\nabla/i$ is the linear momentum operator, and $\boldsymbol{A}$ is the vector potential of the light of amplitude $A$ and the polarization vector $\epsilon$. The term appearing in the numerator of Eq. 1, $\langle\psi_c| H_{eR} |\psi_v\rangle$ describes the transition from a state in the valence band to a state in the conduction band and can be written as [82]

$$\mathcal{V}_{cv} = \frac{\hbar e}{icm} \boldsymbol{A}. \langle\psi_c(\boldsymbol{k}_c)| \nabla |\psi_v(\boldsymbol{k}_v)\rangle \tag{3}$$

where, $\boldsymbol{k}_v$ and $\boldsymbol{k}_c$ are the wave vectors associated with valence and conduction band state, respectively. Under the assumption that the photon wave vector is negligible as compared to the electronic wave vector (dipole approximation) [83], the vector potential $\boldsymbol{A}$ can be separated from the expectation value. The term $P_{cv} = \boldsymbol{A}\cdot\langle\psi_c(\boldsymbol{k}_c)| \nabla |\psi_v(\boldsymbol{k}_v)\rangle$ will be generically referred to as the "optical matrix element" in the subsequent discussion.

In Sec. III, we briefly discuss the continuum model of tBLG by including various effects, such as, corrugation [74, 84–88], doping dependent Hartree interactions [84–86, 89, 90] and particle-hole (ph) asymmetry [91, 92]. We identify the perturbed part of the Hamiltonian after employing light-matter interaction. The eigenvectors of the unperturbed Hamiltonian will be treated as the eigenfunctions appearing in $\langle\psi_c(\boldsymbol{k}_c)| H_P |\psi_v(\boldsymbol{k}_v)\rangle$.

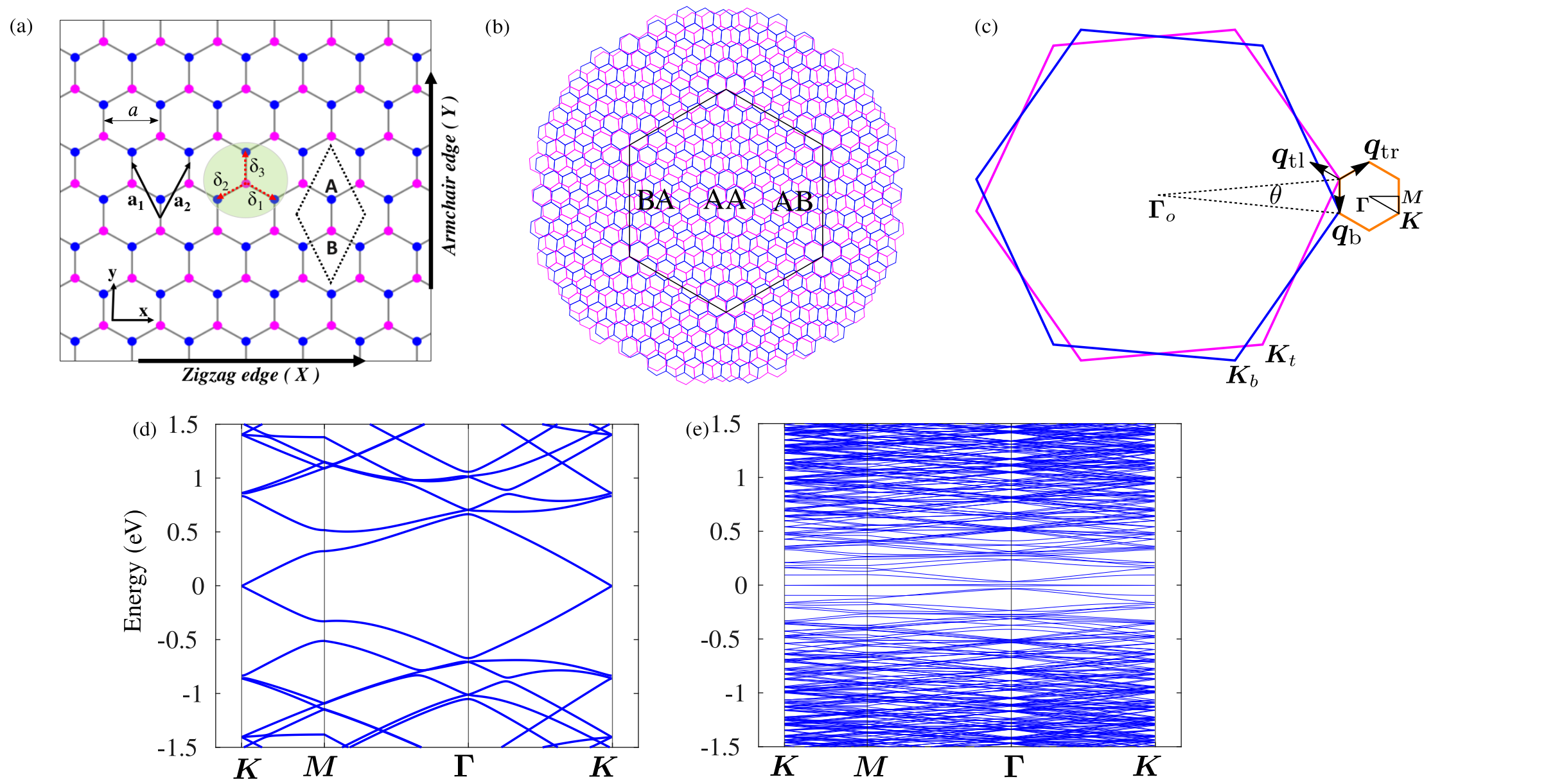

FIG. 2: (a) The real space honeycomb lattice comprising of two carbon atoms $A$ and $B$ with $\mathbf{a}_1$ and $\mathbf{a}_2$ marking the primitive lattice vectors. The black-dotted rhombus marks the unit cell. The first nearest neighbor atoms are highlighted in green circle with vectors $\boldsymbol{\delta}_i$ $(i = 1, 2, 3)$ connecting them. (b) The real space lattice of twisted bilayer graphene at an arbitrary twist angle $\theta$. The black hexagon marks the Moiré unit cell. (c) The pink and blue hexagons represent the rotated BZ, where $\boldsymbol{K}_t$ and $\boldsymbol{K}_b$ represent the Dirac points of top and bottom layer, respectively, separated by $\frac{8\pi}{3a}\sin\left(\frac{\theta}{2}\right)$. A small red hexagon outlines the MBZ. (d) Band structure along the high symmetry path (HSP) of the MBZ for $\theta = 5°$ and (e) $\theta = 1.05°$, respectively, with $w_o = 0.0797$ eV and $w_1 = 0.0975$ eV. Values adapted from Ref. [74].

## III. INTERACTION OF THE PHOTON WITH TWISTED BILAYER GRAPHENE

Fig. 2 (b) represents a two-dimensional (top) view of tBLG with two graphene layers marked in pink (rotated by $\theta/2$) and blue (rotated by $-\theta/2$) color, such that the angle of misorientation between the two layers is $\theta$ with the $AA$ and $AB$-rich regions marked. The black hexagon outlines the Moiré unit cell. Fig. 2 (c) shows the corresponding rotated Brillouin zones where $K_t$ and $K_b$ mark the Dirac points of the top and bottom layers, separated by the vectors $\mathbf{q}_\mathrm{b}$, $\mathbf{q}_\mathrm{tr}$ and $\mathbf{q}_\mathrm{tl}$ as shown. They represent the momentum transfer corresponding to the three inter-layer hopping processes shown in Fig. 2 (c). By considering the effects of corrugation, doping-dependent Hartree interactions and ph asymmetry, the real-space model Hamiltonian for tBLG is [29, 30, 74, 91, 92]

$$H(\boldsymbol{r}) = \begin{pmatrix} h\left(\theta/2\right) + V_H(\boldsymbol{r}) & T'(\boldsymbol{r}) \\ T'^{\dagger}(\boldsymbol{r}) & h\left(-\theta/2\right) + V_H(\boldsymbol{r}) \end{pmatrix} \tag{4}$$

where, $h\left(\theta\right) = -i\hbar\ v_F \boldsymbol{\sigma}_\theta \cdot \boldsymbol{\nabla}$ and $\boldsymbol{\sigma}_\theta = e^{i\sigma_z\ \theta/2}\left(\sigma_x, \sigma_y\right)e^{-i\sigma_z\ \theta/2}$ are the rotated Pauli matrices. The electronic band structure of doped tBLG changes significantly when electron-electron interactions are included *via* self-consistent Hartree calculations [84–86, 89, 90]. For our purpose, we employ the parametrization provided by Goodwin *et al.* [85], according to whom the doping and twist angle-dependent Hartree potential energy is described by

$$V_H(\boldsymbol{r}) \approx \mathcal{V}_\theta \sum_{j=1}^{3} \cos\left(\boldsymbol{G}_j.\boldsymbol{r}\right) \tag{5}$$

where, $\mathcal{V}_\theta = V(\theta)\left[\nu - \nu_o(\theta)\right]$. The quantity $\nu_o(\theta)$ represents the doping level at which the Hartree potential vanishes, $V(\theta)$ is a twist angle- dependent energy parameter, $\nu$ is the filling factor, and $\boldsymbol{G}_j$ denotes the three reciprocal lattice vectors that are used to describe the out-of-plane corrugation of tBLG [74]. This equation has a form similar to the continuum model provided in Ref. [89]. The spatially dependent interlayer tunneling matrices $T'(\boldsymbol{r})$ appearing as off-diagonal elements in Eq. 4 forms a smooth moiré potential [29]

$$T'(\boldsymbol{r}) = \sum_{j=1}^{3} e^{-i\mathbf{q}_j\cdot\boldsymbol{r}} T'_j \tag{6}$$

Considering the initial configuration as the $AA-$ stacking, the $T'_j$ matrices are given by [91–93]

$$T'_j = \left(w_o\sigma_o + iw_3\sigma_z\right) + w_1\left[\sigma_x\cos\left(\phi'\right) + \sigma_y\sin\left(\phi'\right)\right] \tag{7}$$

Here, $\phi' = 2\pi(j-1)/3$ and $\boldsymbol{\sigma}$ are the Pauli matrices. The $w_o$ term contributes to the diagonal elements and represents the interlayer coupling between the A(B) sublattice of the top layer and the A(B) sublattice of the bottom layer. The $w_1$ term only contributes to the off-diagonal elements and is thus associated with the interlayer coupling between the A(B) sublattice of the top layer and B(A) sublattice of the bottom layer. $w_3$ is defined as the interlayer contact coupling and as shown by Kang *et al.* [92], it accounts for the dominant source of the non-negligible ph asymmetry in the model of Ref. [14]. Setting $w_o = w_1$, $w_3 = 0$ and keeping $\mathbf{q}_\mathrm{j}$ in the first shell recover the tunneling matrices of the original BM continuum model [29]. A minor contribution arises from the gradient coupling $\lambda$ which has therefore not been taken into account for the purpose of our calculations in this paper. We show in Appendix. B that the Rayleigh scattering process remains unaffected by the inclusion of ph asymmetry and doping-dependent Hartree interactions in the system. The corrugation effect substantially alter the electron band structure near the $\mathbf{\Gamma}$ point of the MBZ and can be included by setting $w_o \neq w_1$ in Eq. 7.

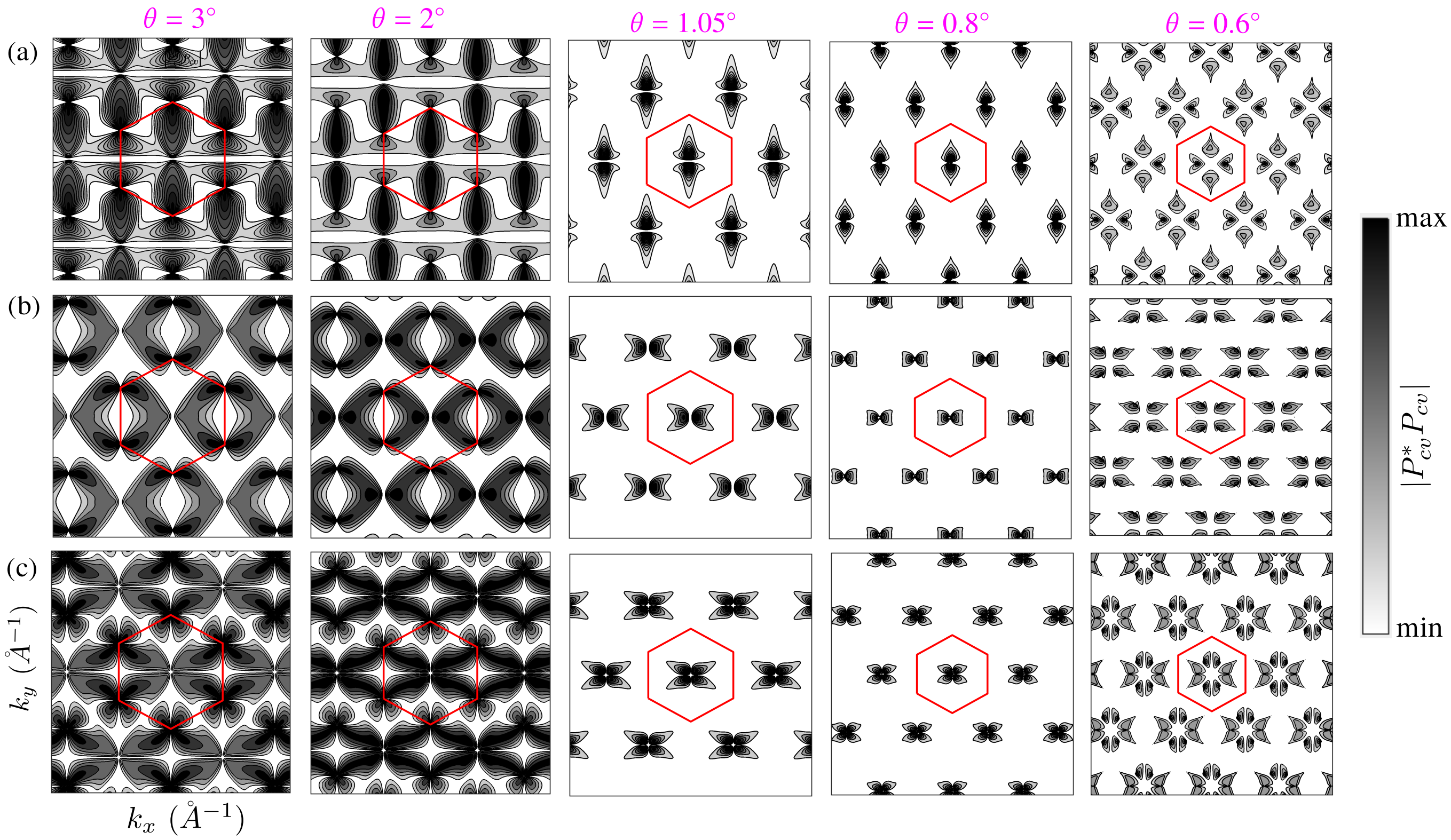

FIG. 3: Contour plots for the optical matrix elements $P_{cv}$ as a function of $k_x$ and $k_y$ in the entire MBZ corresponding to incoming and outgoing polarization directions (a) zigzag: zigzag ($XX$) (b) armchair: armchair ($YY$) and (c) zigzag: armchair ($XY$), respectively for various twist angles. The Red solid line outlines the corresponding hexagonal MBZ.

By switching on the light-matter interaction, the operator $\boldsymbol{\nabla}$ is replaced by $\boldsymbol{\nabla} \rightarrow \boldsymbol{\nabla} + ie\boldsymbol{A}/\hbar$ [94–96], where $\boldsymbol{A}$ is the vector potential of incident light. Under this interaction, the Hamiltonian of tBLG in the simplest limit in which the momentum-space lattice is truncated at the first honeycomb shell is

$$H(\boldsymbol{k}) = \begin{pmatrix} h_{\boldsymbol{k}'}\left(\frac{\theta}{2}\right) & T'_\mathrm{b} & T'_\mathrm{tr} & T'_\mathrm{tl} \\ T'^{\dagger}_\mathrm{b} & h_{\boldsymbol{k}'_\mathrm{b}}\left(-\frac{\theta}{2}\right) & \mathcal{V}_\theta \mathcal{I} & \mathcal{V}_\theta \mathcal{I} \\ T'^{\dagger}_\mathrm{tr} & \mathcal{V}_\theta \mathcal{I} & h_{\boldsymbol{k}'_\mathrm{tr}}\left(-\frac{\theta}{2}\right) & \mathcal{V}_\theta \mathcal{I} \\ T'^{\dagger}_\mathrm{tl} & \mathcal{V}_\theta \mathcal{I} & \mathcal{V}_\theta \mathcal{I} & h_{\boldsymbol{k}'_\mathrm{tl}}\left(-\frac{\theta}{2}\right) \end{pmatrix} \quad (8)$$

where $\boldsymbol{k}'_\mathrm{j} = \boldsymbol{k} + \mathbf{q}_\mathrm{j} + e\ \boldsymbol{A}/\hbar$ with (j=b, tr, tl). Extracting the $\mathbf{A}$-dependent part of Eq. 8 gives

$$H_P = \begin{pmatrix} M_1 & 0 & 0 & 0 \\ 0 & M_2 & 0 & 0 \\ 0 & 0 & M_2 & 0 \\ 0 & 0 & 0 & M_2 \end{pmatrix} \quad (9)$$

as the perturbation. In Eq. 9, $M_j$ is a square matrix of dimensions $2 \times 2$ which has has a dependency on the $x$ and $y$ components of the vector potential $\boldsymbol{A}$ of the incident light as

$$M_j = c_1 \begin{pmatrix} 0 & (A_x + i\ A_y)e^{i(-1)^j\theta/2} \\ (A_x - i\ A_y)e^{i(-1)^{j+1}\theta/2} & 0 \end{pmatrix} (10)$$

where, $c_1 = \hbar v_f e$ and $j = 1(2)$ for the graphene layer rotated by $\theta/2(-\theta/2)$, respectively. The diagonal terms of Eq. 8 represent the Hamiltonian for an isolated rotated graphene layer of the

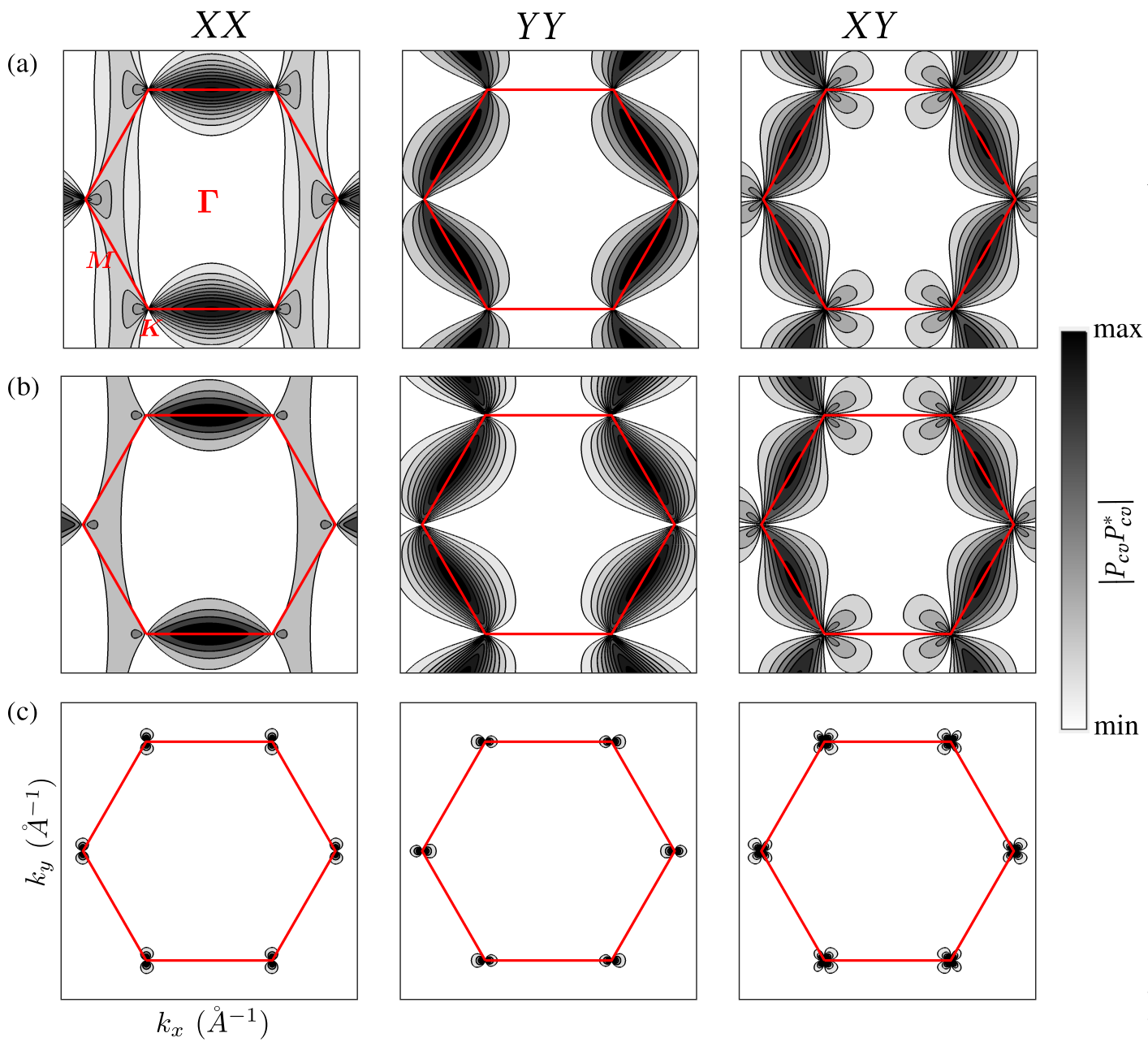

FIG. 4: Contour plots of the optical matrix elements as a function of $k_x$ and $k_y$ in the entire Brillouin zone (measured in . units) for (a) SLG (b) AB-BLG corresponding to the transition between direct bands $v_\alpha \longrightarrow c_\beta$ ($\alpha = \beta$) and (c) AB-stacked BLG for transition between crossing bands $v_\alpha \longrightarrow c_\beta$ ($\alpha \neq \beta$) for incoming and outgoing polarization directions as zigzag: zigzag ($XX$), armchair: armchair ($YY$) and zigzag: armchair ($XY$). The Red solid lines outline the corresponding hexagonal Brillouin zone.

form

$$h_{\boldsymbol{k}}(\theta) = -\hbar v_f |k| \begin{pmatrix} 0 & e^{i(\theta_{\boldsymbol{k}}-\theta)} \\ e^{-i(\theta_{\boldsymbol{k}}-\theta)} & 0 \end{pmatrix} \tag{11}$$

where, $v_f$ is the Fermi velocity of SLG, $\boldsymbol{k}$ is momentum measured from the layer's Dirac point, and $\theta_k = \tan^{-1}\left(\frac{k_y}{k_x}\right)$ is the momentum orientation relative to the x-axis. The off-diagonal matrices are defined by Eq. 7.

In order to calculate the optical matrix elements and Rayleigh intensity, the eigenvectors of the unperturbed, i.e, the **A**-independent part of the Hamiltonian given by Eq. 8 will be treated as the eigenfunctions appearing in $\langle\psi_c(\boldsymbol{k}_c)|\, H_P\, |\psi_v(\boldsymbol{k}_v)\rangle$, where the perturbed Hamiltonian $H_P$, is given by Eq. 9.

In order to select the optimum model for our calculations, we provide a step-wise modification to the pristine BM model by including corrugation [74, 84–87], doping-dependent Hartree interactions [84–86, 89, 90] and ph asymmetry [91, 92] in Appendix B, and infer that the latter two have a negligible effect on the Rayleigh response of twisted bilayer graphene. Therefore, in Sec. IV, we present our results based on the model that incorporates the corrugation effects in the pristine BM model. For reference, the band structure of two sample twist angles, $\theta = 5°$ and $1.05°$ are presented in Figs. 2 (d) and (e), respectively, for $|E| \leq 3$ eV , with the highest bands converged up-to $10^{-5}$eV (by diagonalizing the Hamiltonian of dimensions $232 \times 232$ and $1178 \times 1178$, respectively). It is observed that as the twist angle is lowered, the complexity of the manifold of electronic bands increases markedly within the same energy range [29, 32, 97].

## IV. RESULTS

### A. Optical matrix elements

The optical matrix elements describe the transition between a valence band (VB) state and a state in the conduction band (CB) and are given by $\boldsymbol{A}\cdot\langle\psi_c(\boldsymbol{k}_c)|\, \nabla\, |\psi_v(\boldsymbol{k}_v)\rangle$, where $\boldsymbol{k}_c$ and $\boldsymbol{k}_v$ are the wave vectors corresponding to the conduction and valence band, respectively. Depending on the polarizer-analyzer orientation, we have several possible combinations of the incoming and outgoing polarization. For the sake of brevity in our paper, we will focus on two cases of the parallel polarization on the one hand and cross-polarization on the other.

We explicitly evaluate the optical matrix elements for tBLG by considering transitions between all the possible pairs of valence and conduction bands within an energy range $|E| \leq 3$ eV. We represent the contour plots of these matrix elements as a function of $k_x$ and $k_y$ in the entire MBZ for incoming and outgoing polarization directions as $XX$, $YY$ and $XY$, respectively, in Fig. 3 (a-c) by considering a transition between the lowest pair of valence and conduction band for various twist angles. We find that as the twist angle $\theta$ reduces from 3° to 0.8°, the dominant regions of the MBZ contributing to the matrix elements shift away from $\boldsymbol{K}$ point towards the centre of the MBZ, *i.e.*, the $\Gamma$ point. At very small twist angles, *e.g.*, $\theta = 0.6°$, the location of these wavevectors start shifting from the $\Gamma$ point to the $\boldsymbol{M}$ point of the MBZ.

For comparison, we provide the contour plots of the optical matrix elements for SLG and AB-BLG [98–100] in Figs. 4 (a) and (b-c), respectively. We observe that for direct transitions, *i.e.*, $v_1 \longrightarrow c_1$ or $v_2 \longrightarrow c_2$ in AB-BLG, these matrix elements vanish at the Dirac points (Refer Fig. 4 (b))-in contrast to the case of SLG (Fig. 4 (a)). However, for the transitions involving the crossing pair of bands (Refer Fig. 4 (c)), *i.e.*, $v_1 \longrightarrow c_2$ or $v_2 \longrightarrow c_1$, optical matrix elements show a maximum precisely at the Dirac points. Unlike tBLG, for SLG or AB-BLG, there is only a negligible contribution to these matrix elements from the centre of the Brillouin zone $\Gamma$. The reason lies in the well-established band structure of SLG and AB-BLG. In both cases, there is an energy gap of $\approx 16$ eV [77] at the $\Gamma$ point which is inaccessible by optical experiments. For a more easily accessible optical laser energy $\approx 2$ eV, we are forced to confine ourselves near the $\boldsymbol{K}$ point. The band structure of tBLG represented in Fig. 2 (d-e) suggests that it is reasonably possible to cover the entire MBZ with the same optical laser energy source used for SLG or AB-BLG.

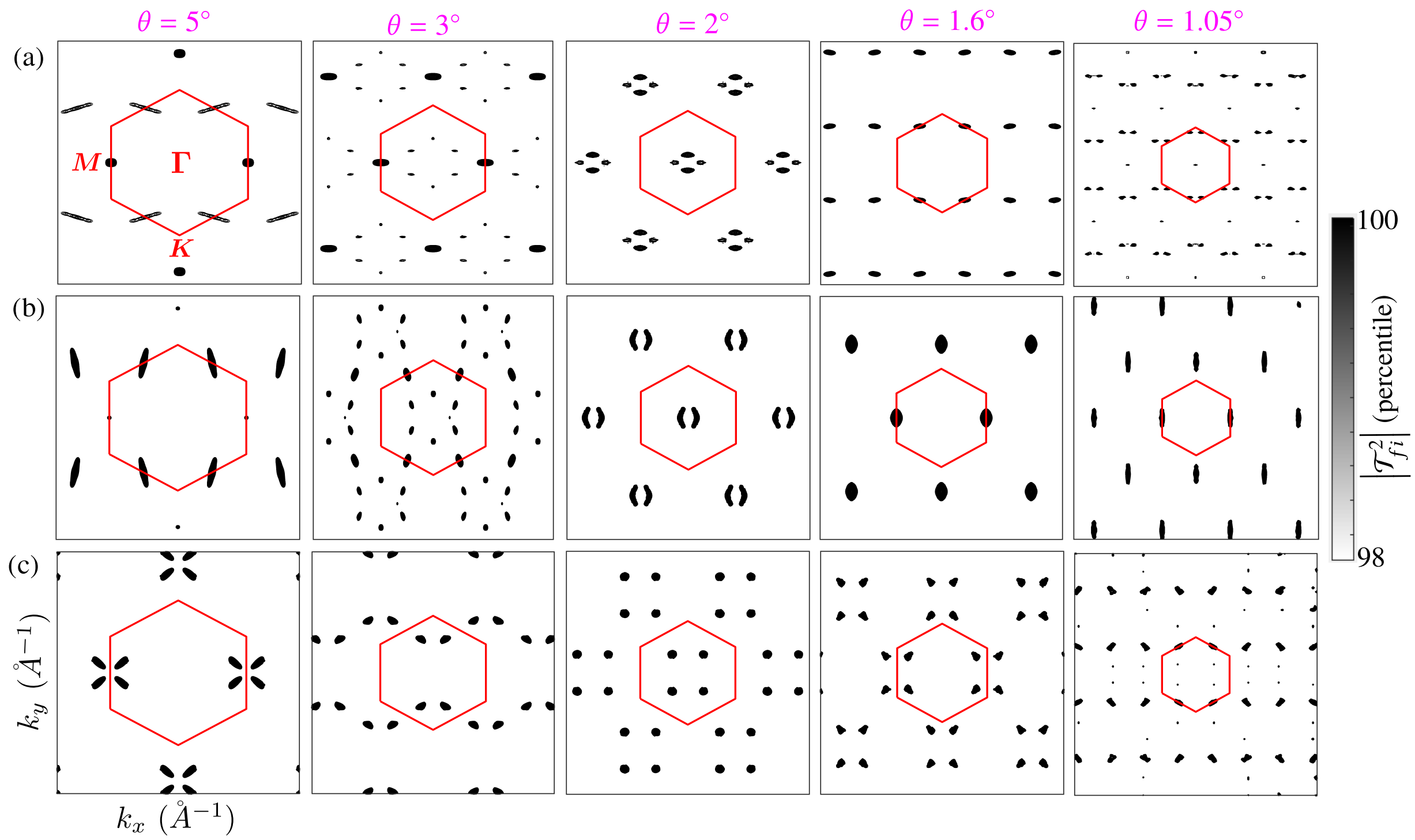

FIG. 5: Contour plots showing the strength of transition matrix $|\mathcal{T}_{fi}|^2$ (at laser energy $E_l = 2.2$ eV) as a function of $k_x$ and $k_y$ in the entire MBZ corresponding to (a) $XX$ (b) $YY$and (c) $XY$ polarization cases. The Red solid line outlines the corresponding hexagonal MBZ.

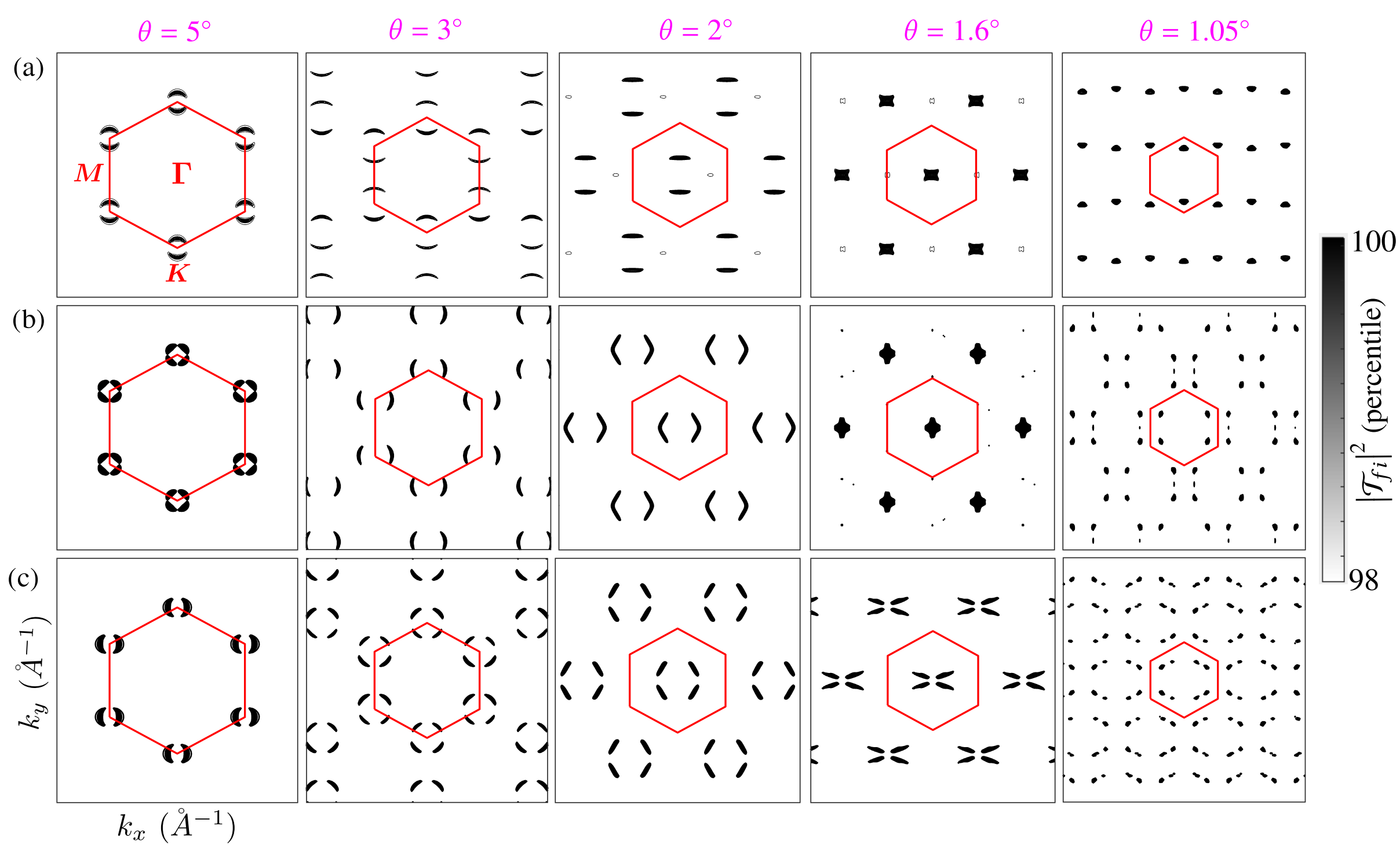

FIG. 6: Contour plots showing the strength of transition matrix $|\mathcal{T}_{fi}|^2$ (at laser energy $E_l = 0.2$ eV) as a function of $k_x$ and $k_y$ in the entire MBZ corresponding to (a) $XX$ (b) $YY$and (c) $XY$ polarization cases. The Red solid line outlines the corresponding hexagonal MBZ.

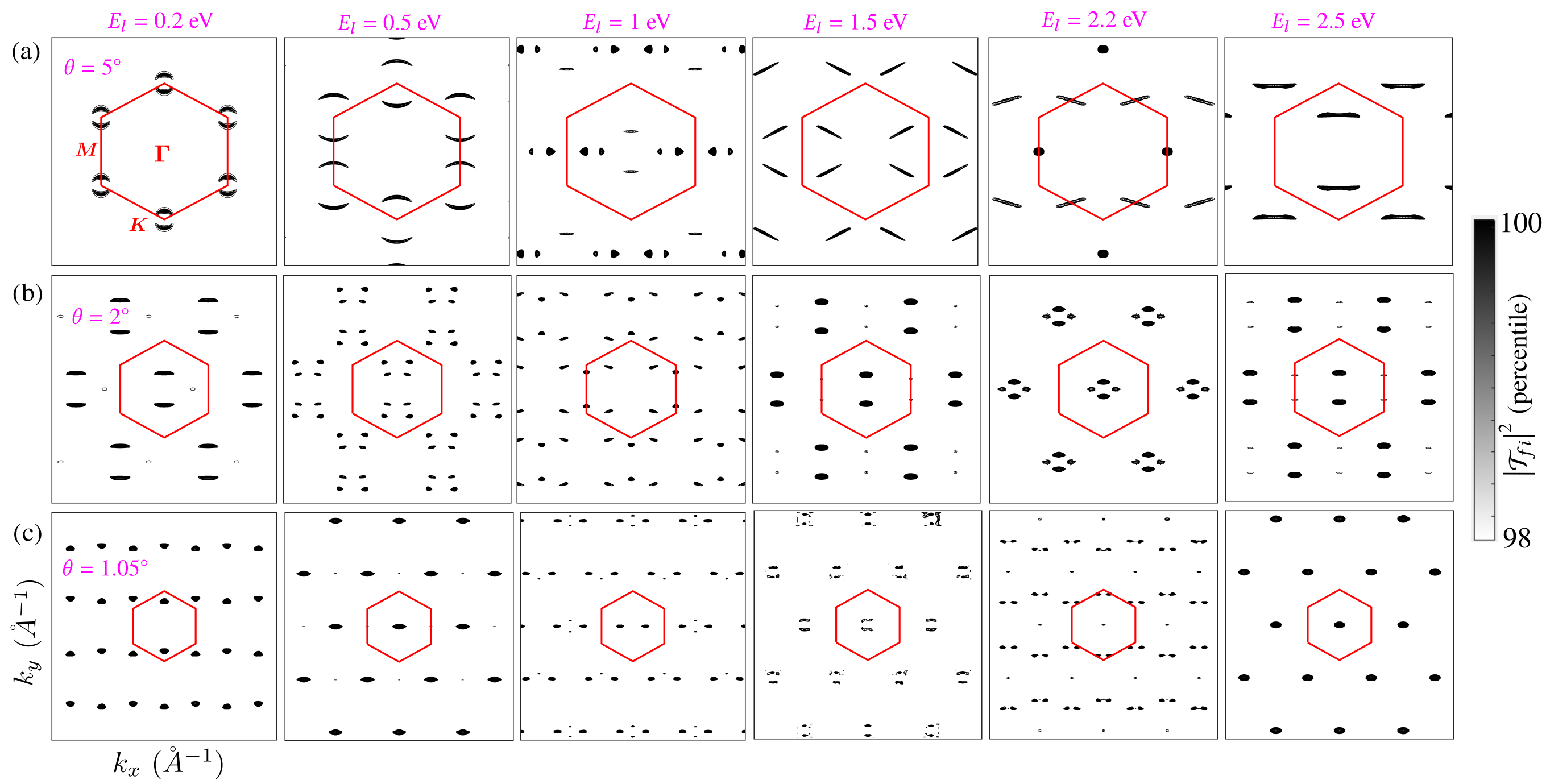

FIG. 7: Contour plots for $\left|\mathcal{T}_{fi}\right|^2$ for varying laser energies as a function of $k_x$ and $k_y$ in the entire MBZ corresponding to parallel polarization ($XX$) for twist angle (a) $\theta = 5°$, (b) $\theta = 2°$, and (c) $\theta = 1.05°$. The Red solid line outlines the corresponding hexagonal MBZ.

### B. Transition matrix: *identifying the resonant region*

When an electron makes a transition from an initial state to a final state by interacting with a photon of energy $E_l$, the probability of transition is maximum when $\left|E_l - (E_f - E_i)\right| \approx 0$, *i.e.*, the energy gap between the initial and final state is in *resonance* with the incoming laser energy $E_l$. $E_f$ and $E_i$ represent the energies of the final and initial transition states, respectively. For Rayleigh scattering, the transition matrix is defined by the following equation [73, 78]

$$\mathcal{T}_{fi} = \sum_m \frac{\langle\psi_f|\, H_{eR}\, |\psi_m\rangle\, \langle\psi_m|\, H_{eR}\, |\psi_i\rangle}{(E_i + E_l - (E_m + \imath\gamma))} \tag{12}$$

The summation is carried over all the possible intermediate states $m$. $|\psi_i\rangle$, $|\psi_m\rangle$ and $|\psi_f\rangle$ represent the electronic states for initial, intermediate and final transition states with energies $E_i$, $E_m$ and $E_f$, respectively. Due to the unavailability of the dependence of broadening parameter $\gamma$ at low laser energies in the literature, we have assumed that the broadening parameter $\gamma$ is $\frac{1}{20}th$ of the laser energy $E_l$. The Rayleigh intensity is proportional to the absolute value of transition matrix $\left|\mathcal{T}_{fi}\right|^2$ [73] and the contour plots for $\left|\mathcal{T}_{fi}\right|^2$ as a function of $k_x$ and $k_y$ in the entire MBZ at laser energies 2.2 eV and 0.2 eV are shown in Figs. 5 and 6, respectively. In order to identify the most dominant region of the Brillouin zone contributing to the Rayleigh scattering, we have restricted our data range to the top 2 percentile. Parts (a), (b), and (c) of these Figs. correspond to the incoming and outgoing polarization directions $XX$, $YY$ and $XY$, respectively. The laser energy $\approx$ 2.2 eV marks the onset of visible spectra and is easily accessible for experimental purposes. We observe that the location of the dominant wave vectors that contribute to Rayleigh scattering is highly dependent on the twist angle $\theta$ as well as the incoming laser energy $E_l$. A detailed discussion of this dependency is presented in the following subsections.

#### 1. *Incoming laser energy $E_l$ : fixed, twist angle $\theta$ : varied*

Fig. 5 shows the contour plots of $\left|\mathcal{T}_{fi}\right|^2$ as a function of $k_x$ and $k_y$ at a fixed incoming laser energy $E_l$ = 2.2 eV for parallel and cross-polarization. For the former case, we observe that at higher twist angles, such as, $\theta = 5°$, the wave vectors near the $\boldsymbol{M}$ point of the MBZ are responsible for the dominant contribution to the Rayleigh scattering process. This region gradually shifts to the centre of MBZ for twist angle $\theta \approx 2°$. At $\theta = 1.6°$, the resonant region is observed to make a transition from $\boldsymbol{\Gamma}$ to $\boldsymbol{M}$ point of the MBZ. For smaller angles, like $\theta = 1.05°$, this region shifts towards the $\boldsymbol{K}$ point of MBZ. Similar behavior is observed for the case of cross-polarization.

Fig. 6 shows the contour plots of $\left|\mathcal{T}_{fi}\right|^2$ for a low incoming laser energy $E_l$ = 0.2 eV for parallel and cross-polarization. In this case we see that the higher twist angles like $\theta = 5°$ exhibit similar behavior as that of SLG (refer Appendix A for detail), *i.e.*, the wave vectors giving the dominant contribution are restricted around the region near $\boldsymbol{K}$ point of MBZ. The resonant region is observed to make a gradual shift from $\boldsymbol{K}$ point to the $\boldsymbol{\Gamma}$ point of the MBZ as the twist angle is reduced from 5° to 1.6°. At $\theta = 1.05°$, the major contribution arises from the region around around halfway between the $\boldsymbol{K}$ and $\boldsymbol{M}$ points.

### 2. *twist angle $\theta$ : fixed , Incoming laser energy $E_l$ : varied*

As indicated by the denominator of Eq. 12, we see that the transition matrix $|\mathcal{T}_{fi}|^2$ is dependent on the incoming laser energy $E_l$. In order to see the effect of this dependency, we identified the wave vectors responsible for the maximum contribution to the Rayleigh scattering process for incoming laser energies varying from $E_l = 0.2$ eV to 2.5 eV. Fig. 7 provides a clear indication of the involvement of the different regions of the MBZ in the scattering process as the laser energy is varied from 0.2 eV to 2.5 eV for (a) $\theta = 5°$, (b) $\theta = 2°$, and (c) $\theta = 1.05°$, respectively. For the sake of conciseness, we have presented this result only for the case of parallel polarization ($XX$). Similar behavior is observed for cross-polarization. We observe that for higher twist angles like $\theta = 5°$ (refer Fig. 7 (a)), at low values of laser energies (0.2 eV or 0.5 eV), the regions around the $\boldsymbol{K}$ point of MBZ provide the dominant contribution to the Rayleigh scattering intensity. As the laser energy increases, this region moves away from the $\boldsymbol{K}$ point. Similarly, at the magic angle, *i.e.*, $\theta = 1.05°$ (refer Fig. 7 (c)), a continuous evolution of the resonant region is observed from the $\boldsymbol{K}$ point to $\boldsymbol{\Gamma}$ point as the laser energy is increased from 0.5 eV to 1.5 eV. However, at $E_l = 2.2$ eV, the resonant region shifts back near the $\boldsymbol{K}$ point. For $\theta = 2°$, at the same laser energy, *i.e*, 2.2 eV, these wavevectors stem from region round the $\boldsymbol{\Gamma}$ point. From these observations, we infer that, unlike SLG or AB-BLG, the Rayleigh scattering process is highly dependent on the laser energy and each twist angle is characterized by its unique region of resonance. Our analysis also suggests that the strength of the transition matrix $|\mathcal{T}_{fi}|^2$ for the same twist angle is weaker in the cross polarization case compared to parallel one. *E.g.*, at laser energy $E_l = 2.2$ eV, the strength of transition matrix is ~ 99 times weaker in cross-polarization compared to parallel polarization for twist angle $\theta = 2°$.

## C. The Rayleigh spectra and integrated Rayleigh intensity

Fig. 8 (A) shows the variation of the Rayleigh intensity measured by the spectrometer as a function of the emitted photon energy $E_s$, at the incoming laser energy 2 eV for the model case of (a) constant optical matrix elements (b) parallel (XX or YY), and (c) cross- polarization (XY), respectively. The maximum intensity value for parallel polarization increases as the twist angle decreases, but for cross-polarization, an opposite trend is observed. The corresponding integrated Rayleigh intensity (plotted as a function of twist angle) is shown in Fig 8 (B). It is observed that when the optical matrix elements are assumed to be constant, the integrated Rayleigh intensity rises exponentially with the lowering of the twist angle. Fig. 8 (B) shows a similar exponential enhancement observed for parallel polarization case, albeit with a reduction in the value of integrated Rayleigh intensity. However, in the case of cross-polarization,

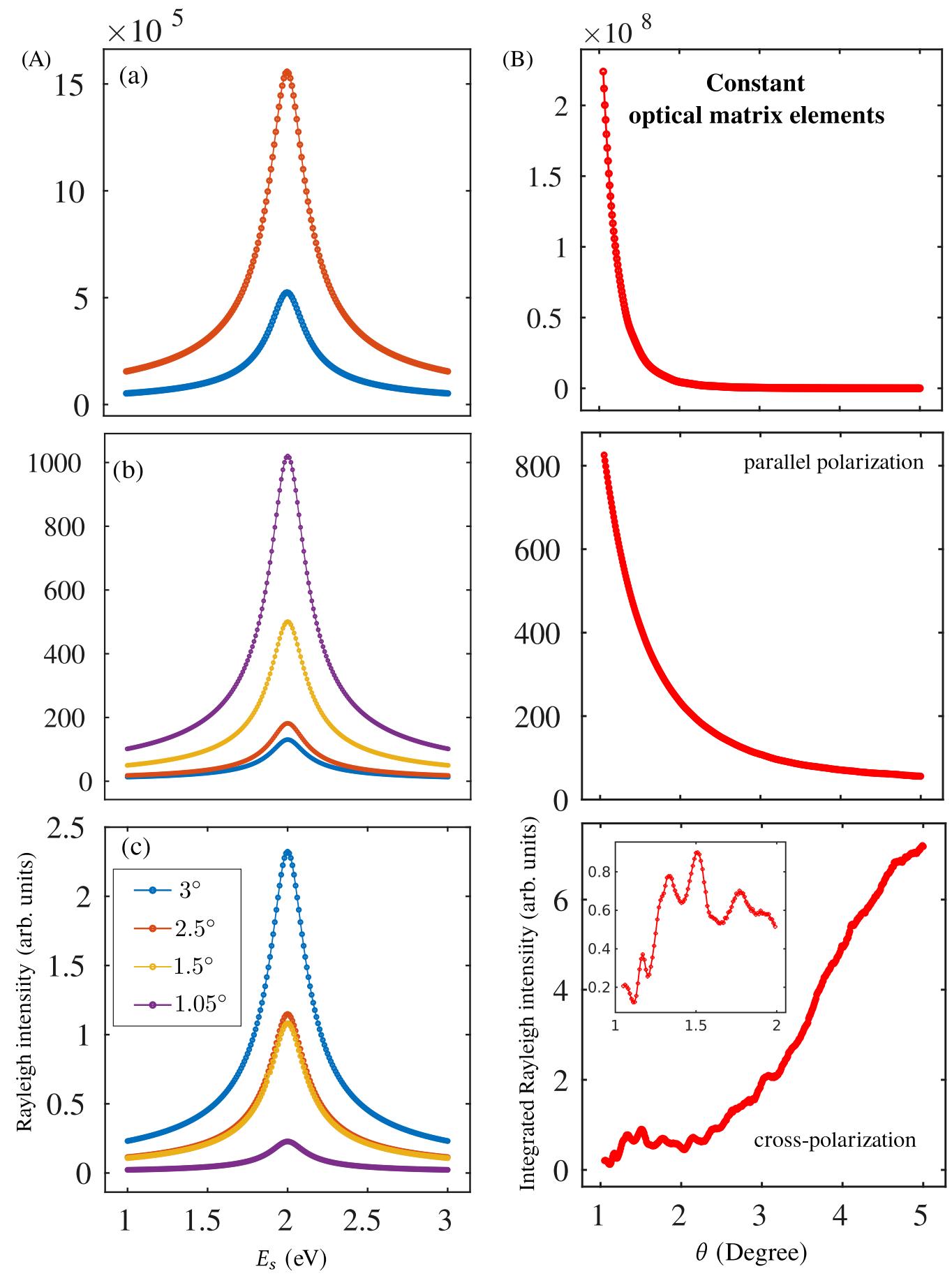

FIG. 8: (A) The Rayleigh intensity as a function of emitted photon's energy for various twist angles. (B) The integrated Rayleigh intensity as a function of twist angle $\theta$ at an incoming laser energy $E_l = 2$ eV for the model case of (a) constant optical matrix elements (b) parallel ($XX$ or $YY$), and (c) cross polarization ($XY$), respectively. The inset shows a zoom-in for small $\theta$.

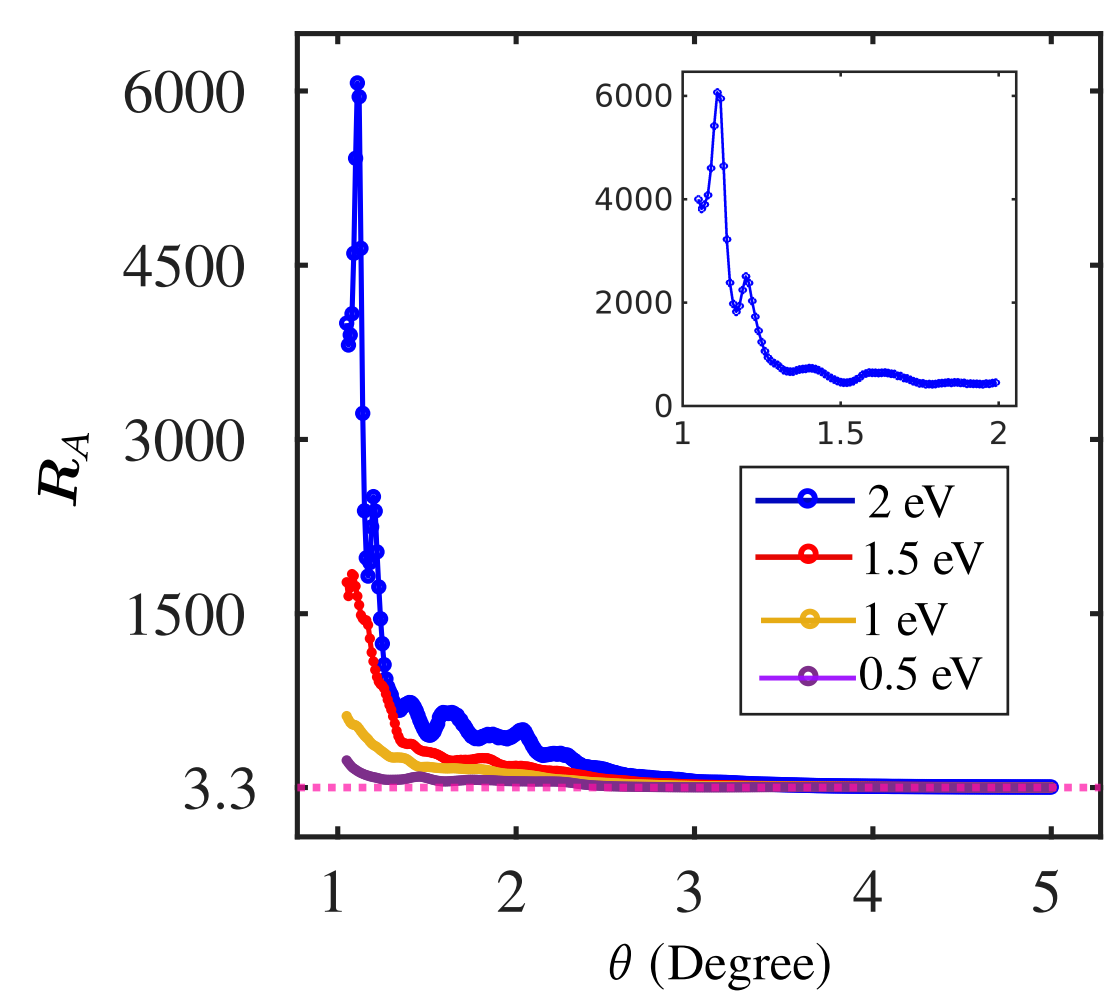

FIG. 9: Variation of $\boldsymbol{R_A}$ with twist angle $\theta$ for various incoming laser energies.

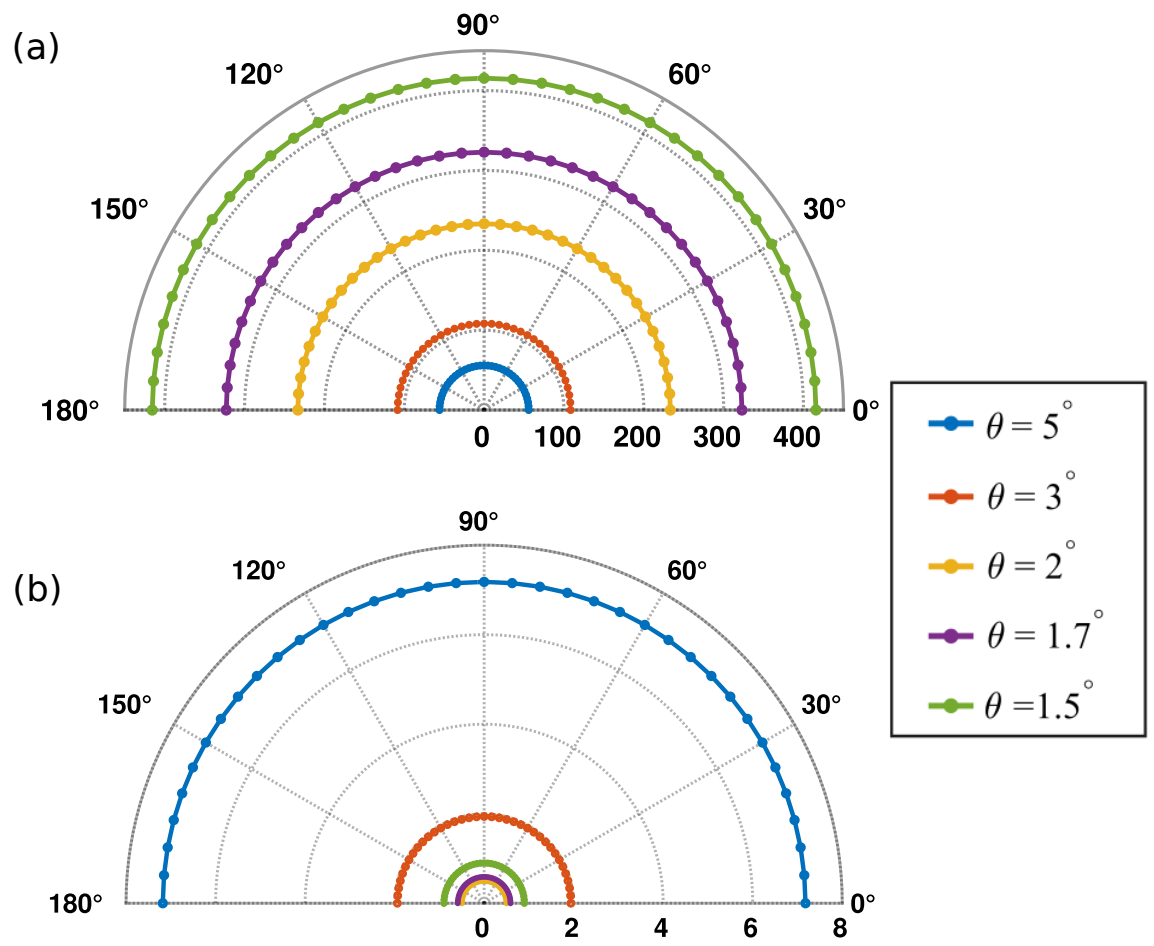

FIG. 10: Variation of the integrated Rayleigh intensity (at $E_l$ = 2 eV) when the polarizer-analyzer orientation is rotated in phase for (a) parallel (b) cross- polarization, respectively.

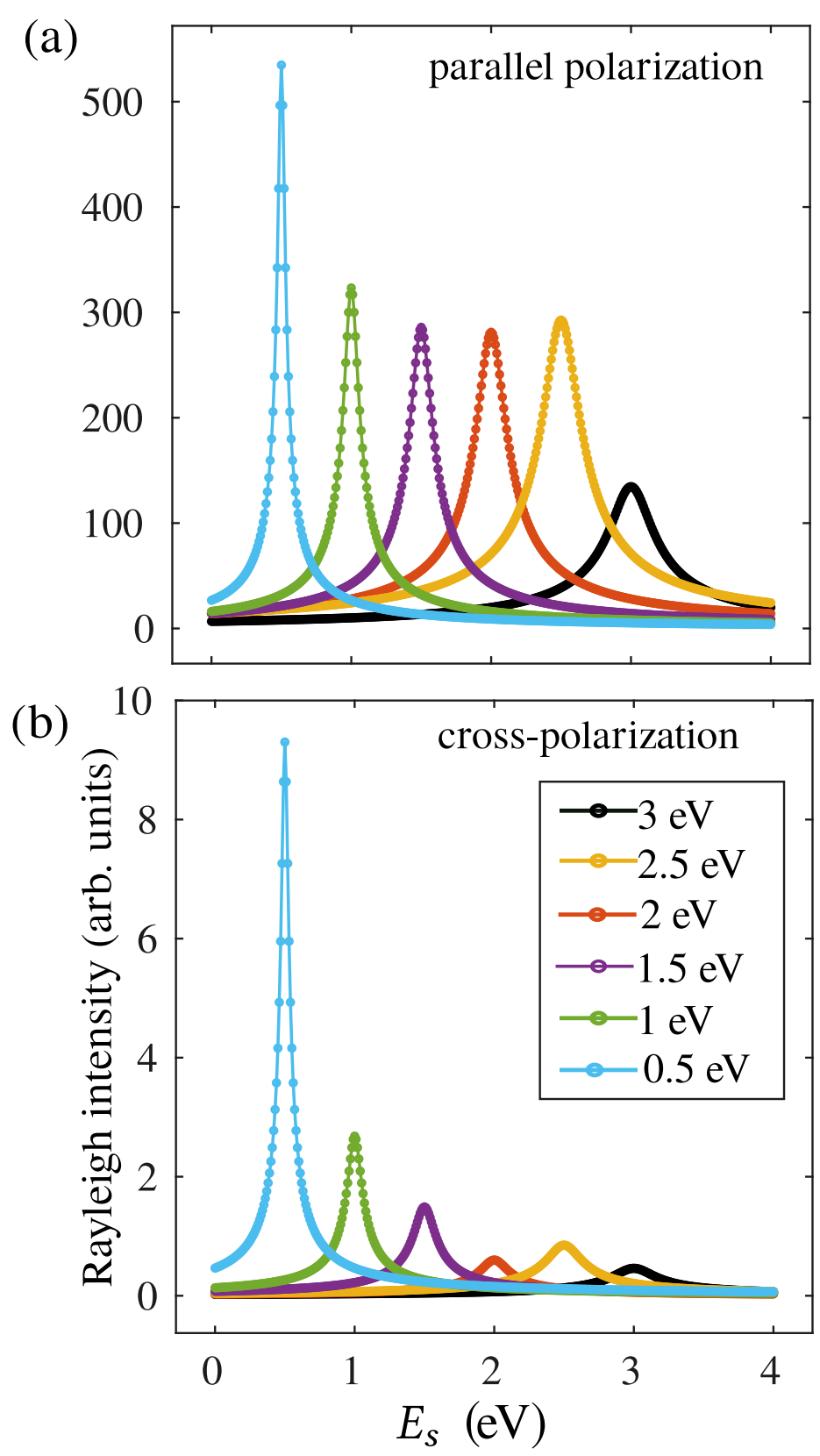

FIG. 11: The Rayleigh intensity (measured in arb. units) for twist angle $\theta = 2°$ at various values of $E_l$ (indicated in the legend) as a function of emitted photon's energy $E_s$ for (a) parallel (b) cross- polarization.

the value of the integrated Rayleigh intensity decreases with the decrease in twist angle (Fig. 8 (B)-c), exhibiting a *hockey stick*-like curve for integrated Rayleigh intensity. A complex behavior is observed at small twist angles with the values strongly diminished compared with the parallel case. Such behavior is suggestive of strong *interference effects* [61] at play, mediated by the incoming and outgoing optical matrix elements for the case of cross-polarization.

The variation of the ratio $\boldsymbol{R}_A$ as the laser energy $E_l$ is increased from 0.5 eV to 2 eV is shown in Fig. 9. Irrespective of the laser energy, the saturation of this ratio $\boldsymbol{R}_A$ to the value 3.33 reminiscent of SLG and AB-BLG graphene is observed at higher twist angles, $\theta \approx 4°$. For a given twist angle, the value of $\boldsymbol{R}_A$ continuously enhances as the laser energy is increased. For low values of twist angles, the ratio $\boldsymbol{R}_A$ is strongly enhanced and is accompanied by oscillations. This is due to the complexity involved in the behavior exhibited by integrated Rayleigh intensity for cross-polarization (Refer Fig. 8 (B)-c) caused due to the aforementioned interplay of interference of the optical matrix elements and complex manifold of bands at such $E_l$.

Fig. 10 exhibits the *isotropic* nature of the integrated Rayleigh intensity when the polarizer and analyzer are rotated in-phase from 0° to 180°. The polar axis marks the angle by which the polarizer and analyzer are rotated (in-phase), and the radial axis marks the magnitude of the integrated Rayleigh intensity (plotted in Fig. 8 (B) in the red solid line).

We plot the Rayleigh intensity as a function of emitted photon energy $E_s$ for different values of incoming laser energies in Fig. 11 for (a) parallel and (b) cross-polarization at a fixed value of twist angle, $\theta = 2°$. For both the cases, we observe that the curve becomes narrower (since the value of broadening parameter $\gamma$ reduces proportionally according to our model) with an increase in the maximum intensity value as the laser energy is decreased from 3 eV to 0.5 eV for parallel polarization. However, a complex behavior in the maximum Rayleigh intensity with the variation in laser energy is observed in the cross-polarization case.

The data represented *via* the red, solid line in Fig. 12 shows the variation of the ratio $\boldsymbol{R}_A$ as the incoming laser energy is varied from 0.1 eV to 3 eV for various twist angles $\theta$. As the twist angle $\theta$ is reduced from 4° to 0.6°, a continuous enhancement is observed in the maximum value of the ratio $\boldsymbol{R}_A$. The peak position is specific to the twist angle. We observe that each twist angle is characterized by a series of unique peaks, which ought to serve as a fingerprint for identifying the twist angle warranting further experimental investigation.

## V. CONCLUSION

In summary, we have carried out a detailed study of polarization-controlled Rayleigh scattering in twisted bilayer graphene (tBLG) and observed the following salient features:

In contrast to SLG or AB-BLG, the dominant wavevectors of the optical matrix elements in tBLG are no longer restricted only to the vicinity of the $\boldsymbol{K}$-point. From Fig. 3, we observe that as the twist angle is reduced from 3° to 0.8°, the location of these wavevectors shifts from $\boldsymbol{K}$ point to the $\Gamma$ of the MBZ. For smaller angles though, *e.g.*, $\theta = 0.6°$, there is a shift observed back to the $\boldsymbol{M}$-point of MBZ. However, in the case of the seminal BM model, the maximum contribution to the optical matrix elements arises exactly from the $\boldsymbol{\Gamma}$ point for $\theta = 1.05°$. By further considering the parametrized continuum models and including the effect of Hartree interactions and ph asymmetry in tBLG, we observe an essentially negligible change in our results.

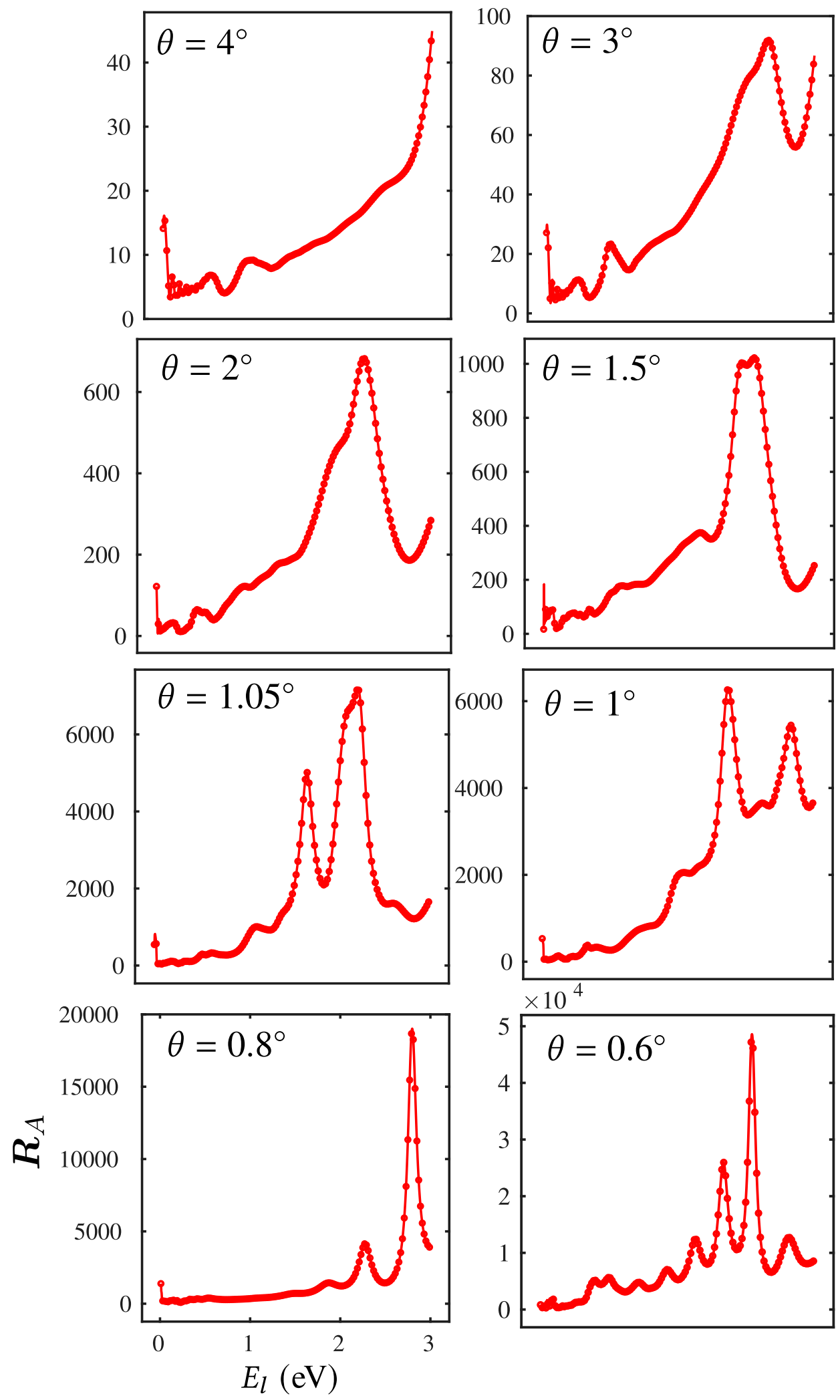

FIG. 12: Variation of $\boldsymbol{R}_A$ with incoming laser energy $E_l$ for different twist angles $\theta$

From Figs. 5, 6 and 7, we observe that the location of wavevectors responsible for the Rayleigh scattering process is highly dependent on the twist angle $\theta$ and incoming laser energy, $E_l$. From the calculations shown in Appendix. B, we also conclude that, irrespective of the incoming laser energy, the location of these wavevectors remain unaffected by the doping dependent Hartree interactions and the ph asymmetry. It can be understood by the fact that doping only modifies the bands nearest to the Fermi level, which has a band gap ($E_g = E_i - E_m$) of few meV. At laser energies used in optical experiments, the transition between these bands plays a negligible role as $E_g \ll E_l$, which may be appreciated by considering the contribution of the denominator of Eq. 12.

The Rayleigh intensity as a function of scattered photon energy is investigated for the model case of (a) constant optical matrix elements (b) parallel, and (c) cross-polarization (refer Fig. 8). We find that the integrated Rayleigh intensity enhances continuously as the twist angle decreases for parallel polarization. In the case of cross-polarization, we observe a *hockey stick*-like curve with complex oscillatory behavior at small twist angles (Refer to the inset of Fig. 8-B (c)). The ratio $\boldsymbol{R}_A$ having a value of $\sim 3.33$ in SLG/AB-BLG is strongly enhanced to $\sim 2500$ for $\theta = 1.2°$ at laser energy $E_l = 2$ eV (showing nearly 3 orders of magnitude increase *viz a viz* SLG and AB-BLG) and begins saturating for $\theta > 4°$ (Refer Fig. 9). It is also observed that for a given twist angle, the value of $\boldsymbol{R}_A$ increases as the laser energy is increased. The integrated intensities are *isotropic* in nature with the in-phase rotation of the polarizer and analyzer from 0 to 180° (Refer Fig. 10).

The Rayleigh intensity spectra (for an arbitrary twist angle $\theta = 2°$) for various values of incoming laser energies are shown in Fig. 11. A broadening of the curve is observed with the increase in laser energy for parallel and cross-polarization. The maximum Rayleigh intensity decreases with the increase in laser energy for the case of parallel polarization, while a complex behavior is observed in the cross-polarization.

The evolution of $\boldsymbol{R}_A$ with the incoming laser energy varying from 0.1 eV to 3 eV for various twist angles is shown in Fig. 12. It suggests that each twist angle is characterized by a series of unique peaks, which serve as a fingerprint for its identification. $\boldsymbol{R}_A$ shows a maximum value at $E_l \approx 3$ eV and $E_l \approx 2.2$ eV for twist angles $\theta = 4°$ and $\theta = 0.6°$, respectively, suggesting that maximum value of $\boldsymbol{R}_A$ is laser energy specific for each twist angle. In Figs. 20 and 21 of Appendix B, we show that the position of these peaks remain unaltered on the inclusion of ph asymmetry and the doping dependent Hartree interactions The $\theta$ dependent quantities appearing in Eq. 5 are taken from Ref. [85].

From our calculations, we conclude that the optical response of tBLG remains unaffected by varying the doping level in the system or by including other interactions, such as ph asymmetry. This is because experimentally, the Rayleigh scattering process is studied under an easily accessible laser energy, like, 2.2 eV, which marks the onset of visible spectra. Any interactions and doping alters the bands nearest to the Fermi-level. The energy gap between these bands is of the order of few meV and hence, the transition between them contributes negligibly to the process.

Our work also represents, to the best of our knowledge [101], the first explicit calculation of the optical matrix elements of twisted bilayer graphene based on the model of Bistritzer and MacDonald [102]. In the preceding sections, we have demonstrated that by switching them off and by contrasting the cases of parallel and cross-polarization, the optical matrix elements are mediators of strong interference effects which are in play in the Rayleigh scattering response of tBLG. Indeed, even at reasonable laser energies, they allow access to parts of the Moiré Brillouin zone of tBLG, which are simply off-limits to those in SLG and AB-BLG. Given that Rayleigh scattering is elementary to all higher-order spectroscopies, *e.g.*, Raman, and two-photon spectroscopies, our results argue that the consideration of the optical matrix elements shall be indispensable to *all* spectroscopic investigations of tBLG, particularly when assessments are made of the origin and relative intensities of its spectral peaks, *e.g.*, the novel peaks observed in the Raman response of tBLG [103, 104]. As regards to the explicit consideration of the incoming and outgoing polarization, we remark that typical laboratory spectroscopic experiments have a preferred polarization not least due to the inherent polarization of the laser source, *e.g.*, a linearly-polarized He−Ne cw-laser. Based on our previous work [63], we also anticipate that our optical-matrix elements-reliant approach would be crucial while interpreting the optical response of tBLG with lowered structural symmetry, *e.g.*, uniaxial and locally distorted tBLG which are extant in the literature [105–108] where polarization-controlled experiments [21] ought to provide a non-isotropic response at a given laser energy $E_l$, therefore providing experimentally facile signatures of the local strain state.

**ACKNOWLEDGMENT**

Disha Arora acknowledges The Department of Science and Technology, Government of India, for supporting this work *via* the INSPIRE fellowship scheme (Ref. No. DST/INSPIRE/03/2017/003017). Deepanshu Aggarwal is supported by the UGC, Government of India Fellowship. The authors would also like to acknowledge Zachary A. H. Goodwin for helpful discussions.

### Appendix A: Analytical expressions for the optical matrix elements

In the first nearest-neighbor approximation, expression for the optical matrix elements describing a transition between a valence and a conduction band state (characterized by wave vectors $\boldsymbol{k}_v$ and $\boldsymbol{k}_c$, respectively) is given by [82, 98, 109]

$$\langle\psi^c(\boldsymbol{k}_c)|\,\nabla\,|\psi^v(\boldsymbol{k}_v)\rangle = -\frac{2\sqrt{3}m_{opt}}{a}\,\mathrm{Re}\left\{c_B^{c*}(\boldsymbol{k})c_A^v(\boldsymbol{k})\sum_l \boldsymbol{r}_A^l exp(-i\boldsymbol{r}_A^l.\boldsymbol{k})\right\} \tag{A1}$$

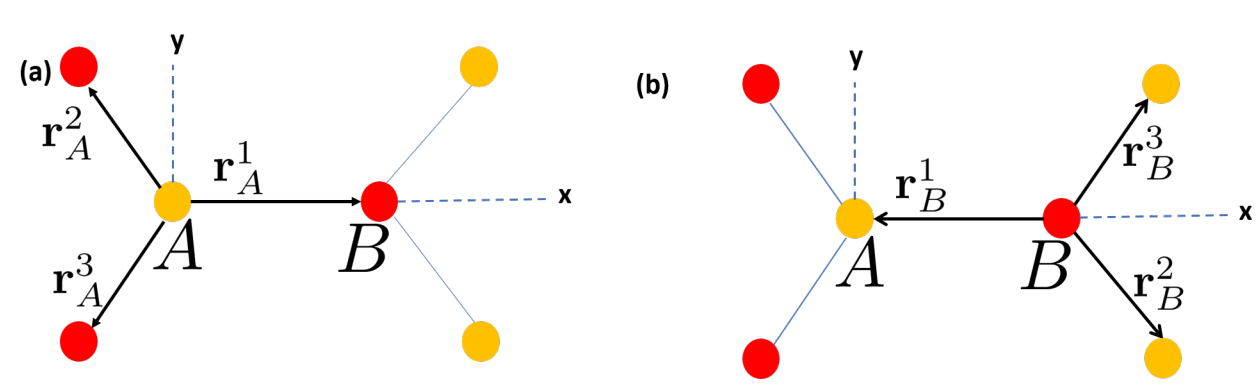

FIG. 13: Vectors connecting the nearest-neighbor atoms from (a) atom $\boldsymbol{A}$ to $\boldsymbol{B}$ and (b) atom $\boldsymbol{B}$ to $\boldsymbol{A}$ *via* vectors $r_A^l$ and $r_B^l$ $(l = 1, 2, 3)$, respectively.

where, $a = 0.246$ nm is the lattice constant of graphene and $m_{opt}$ denotes the constant optical matrix element for the two nearest-neighbor atoms. $r_{\boldsymbol{A}}^l$ $(l = 1, 2, 3)$ are the vectors connecting first nearest-neighbor atoms of graphene shown in Fig. 13. $c_j^i(\boldsymbol{k})$ are the expansion coefficients where the superscript $i$ refers to the valence or conduction band and the subscript $j$ represents the sublattice index.

#### 1. Single-layer graphene

When the vector potential $\boldsymbol{A}$ is polarized along the $X$-direction (*i.e.* along the zigzag-edge of graphene), the optical matrix elements are given as:

$$P_{cv}^x(\boldsymbol{k}) = \frac{\mathrm{Re}[A + B - C]}{2|\exp\left(\frac{iak_y}{\sqrt{3}}\right) + 2\exp\left(\frac{-iak_y}{2\sqrt{3}}\right)\cos(ak_x/2)|} \tag{A2}$$

$$A = \frac{a}{\sqrt{3}}\exp\left(\sqrt{3}iak_y/2\right)2\cos(ak_x/2)$$
$$B = \frac{a}{\sqrt{3}}\exp\left(-\sqrt{3}iak_y/2\right)\cos(ak_x/2)$$
$$C = \frac{a}{2\sqrt{3}}2\cos(ak_x)$$

and for the $Y$-polarized (*i.e.*, along the armchair-edge of graphene) vector potential,

$$P_{cv}^y(\boldsymbol{k}) = \frac{\sqrt{3}\,\mathrm{Re}\,[A1 * B1 * C1]}{2|\exp\left(\frac{iak_y}{\sqrt{3}}\right) + 2\exp\left(\frac{-iak_y}{2\sqrt{3}}\right)\cos(ak_x/2)|} \tag{A3}$$

$$A1 = \exp(-iak_x)$$
$$B1 = -1 + \exp(iak_y)$$
$$C1 = 1 + \exp(iak_x) + \exp\left(ia(\sqrt{3}k_y + k_x)/2\right)$$

Re denotes the real part of [...] and 'a' is the lattice constant of graphene. In the low energy limit, the above two equations reduce to:

$$OM_X(\boldsymbol{k} + \boldsymbol{K}) = \frac{3\,m_{opt}}{2|k|}(k_y, -k_x, 0) \tag{A4}$$

and

$$OM_Y(\boldsymbol{k} + \boldsymbol{K}) = \frac{3\,m_{opt}}{2|k|}(-k_y, k_x, 0) \tag{A5}$$

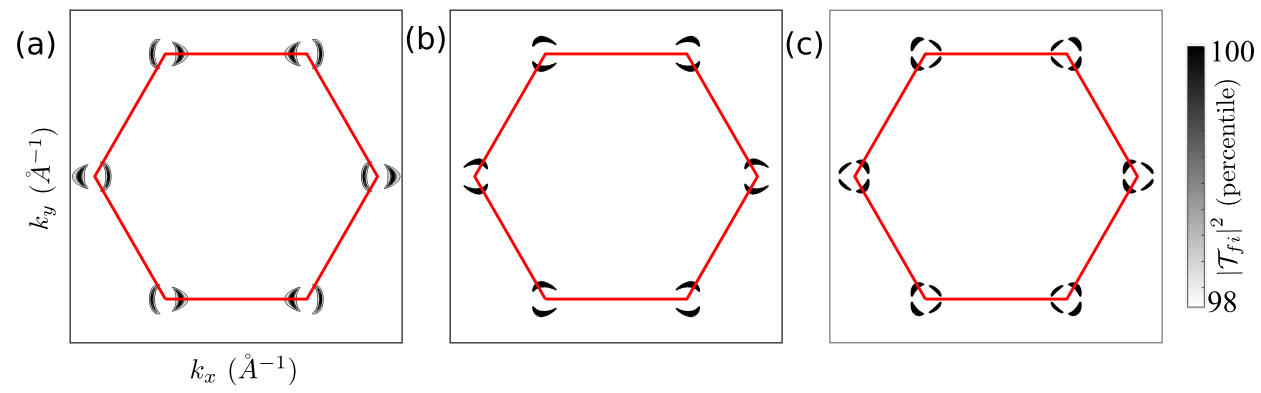

FIG. 14: Contour plots for the strength of transition matrix $|\mathcal{T}_{fi}|^2$ as a function of $k_x$ and $k_y$ for incoming and outgoing polarization as (a) $XX$ (b) $YY$ (c) $XY$ for SLG.

with $m_{opt}$ as a constant (having value 3 nm$^{-1}$) and $|k| = \sqrt{k_x^2 + k_y^2}$.

Depending upon the incoming and outgoing polarization directions, the following expressions are calculated using Eq. A2 and A3.

TABLE I: Expressions to calculate the optical matrix elements for $XX$ (zigzag: zigzag),$YY$ (armchair: armchair) and $XY$ (zigzag: armchair)

| Incoming | Outgoing | Expression |
|---|---|---|
| X | X | $P_{cv}^{*x}(\boldsymbol{k})P_{cv}^{x}(\boldsymbol{k})$ |
| Y | Y | $P_{cv}^{*y}(\boldsymbol{k})P_{cv}^{y}(\boldsymbol{k})$ |
| X | Y | $P_{cv}^{*y}(\boldsymbol{k})P_{cv}^{x}(\boldsymbol{k})$ |

The contour plots for the product of incoming and outgoing optical matrix elements as a function of $k_x$ and $k_y$ in the entire BZ for various polarization cases are plotted in Fig. 4(a). We observe that there is a finite contribution arising only from the wave vectors near the $\boldsymbol{K}$-point. Fig. 14 shows contour plots marking the most resonant region in k-space responsible for the Rayleigh scattering. It shows that for SLG, we are confined to the regions around the $\boldsymbol{K}$-point while the rest of the region of the BZ has a negligible contribution to Rayleigh scattering process. This could be understood from the band-structure of SLG which suggests that the energy gap increases on moving away from the $\boldsymbol{K}$ point making it challenging to access such a large energy gap.

### 2. Bernal-stacked bilayer graphene

The unit cell of AB-BLG comprises of 4 carbon atoms $A_1$, $B_1$ (in layer 1) and $A_2$, $B_2$ (in layer 2) with an inter planar spacing of $d \approx 0.34$ nm. Band structure along high-symmetry path for AB-BLG is shown in Fig. 15 (b). $\gamma_o$ is the tight-binding parameter defining the interaction between two nearest carbon atoms within the same layer and $\gamma_1$ describes the coupling between atoms $A_1$ and $B_2$ on different layers. Blue arrows in Fig. 15 (b) mark the transition between same pair of bands (or direct bands), *i.e.*, $v_1 \longrightarrow c_1$ or $v_2 \longrightarrow c_2$ and green arrows mark the transition between the crossing bands, *i.e.*, $v_1 \longrightarrow c_2$ or $v_2 \longrightarrow c_1$. In the first nearest neighbor tight-binding approximation, the optical matrix elements [100] for transition between interbands are:

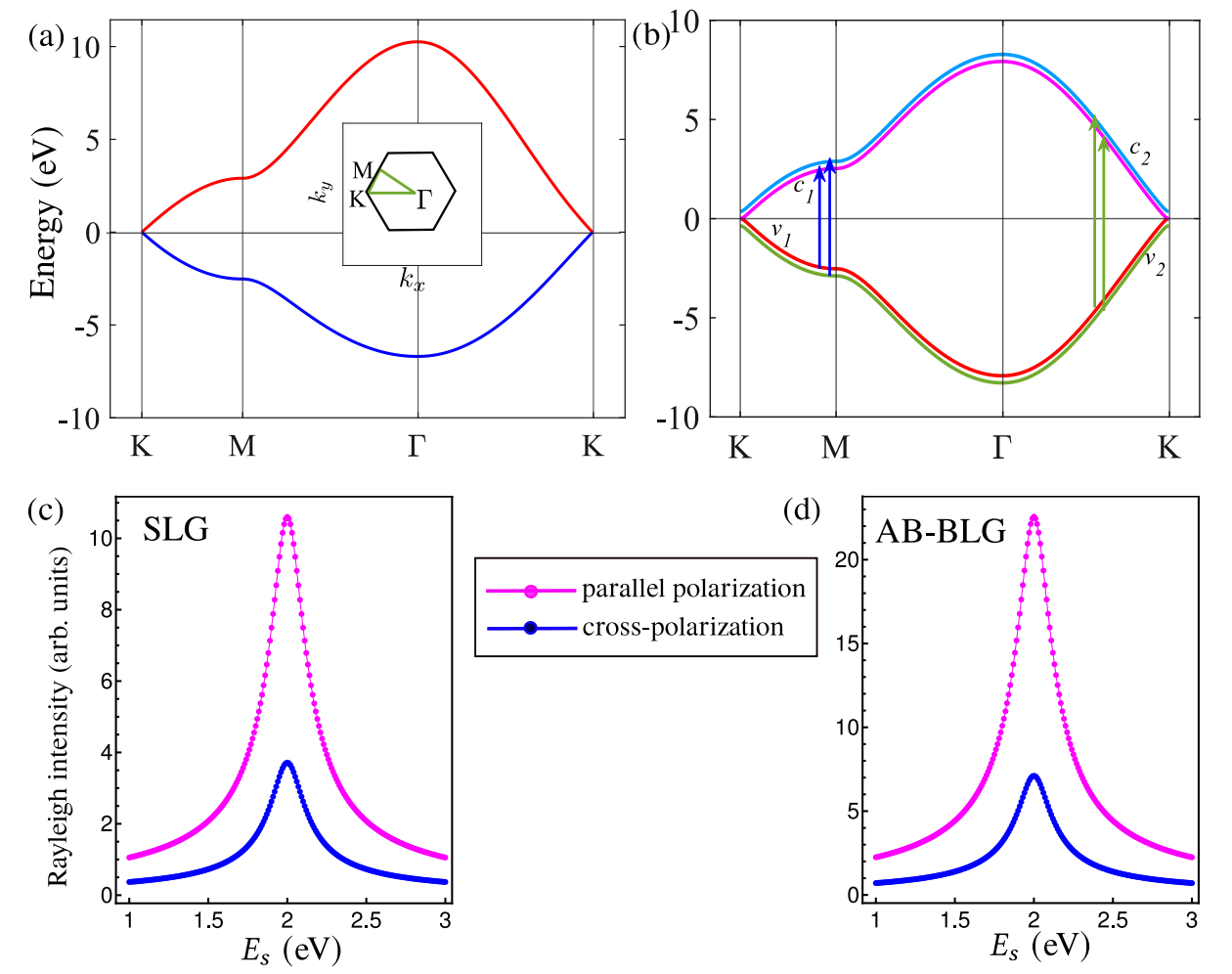

FIG. 15: Band structure along the HSP for (a) SLG and (b) AB-BLG, respectively, with the Rayeligh intensity (as a function of $E_s$) plotted in (c) and (d). The tight binding parameters are set to be $\gamma_o \approx -2.7$ eV and $s_o \approx -0.07$ for SLG. For, AB-BLG, $\gamma_o \approx -2.6$ eV and $\gamma_1 \approx 0.36$ eV (values taken from [100, 110]). It may be noted that $\boldsymbol{R}_A \approx 3.3$ for SLG and BLG, *i.e.*, integrated Rayleigh intensity for cross-polarization is $one - third$ of the integrated Rayleigh intensity for parallel polarization.

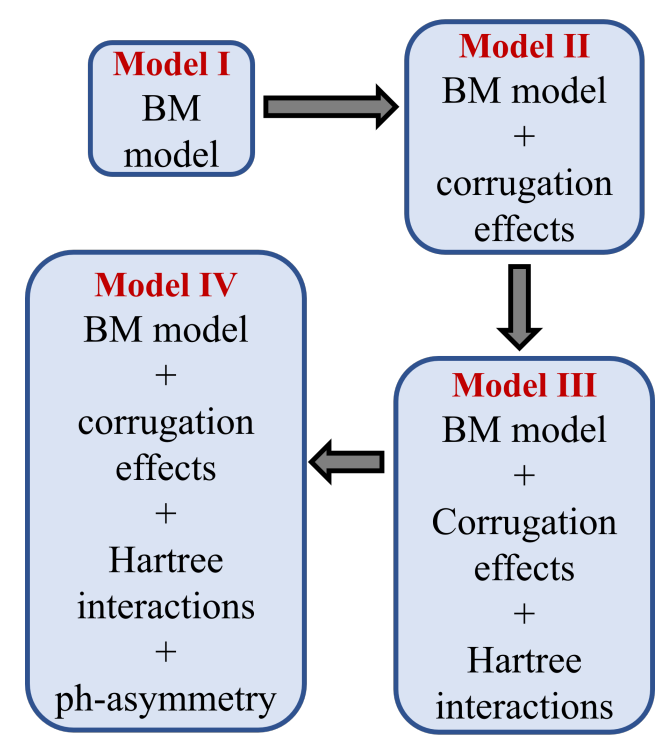

FIG. 16: Step-wise modification to the BM model

$$\mathcal{V}_{\boldsymbol{k}}^{v_\alpha c_\alpha} = \frac{\hbar e}{icm}\boldsymbol{A}.\frac{2m_{opt}\gamma_o}{\sqrt{(4|f(\boldsymbol{k})|^2\gamma_o^2 + \gamma_1^2)}}\text{Re}\ \ [\chi] \tag{A6}$$

and for the transition involving cross bands

$$\mathcal{V}_{\boldsymbol{k}}^{v_\alpha c_\beta} = \frac{\hbar e}{icm}\boldsymbol{A}.\frac{i\, m_{opt}\ \gamma_1}{|f(\boldsymbol{k})|^2\gamma_o}\left(1 + \frac{\gamma_1^2}{|f(\boldsymbol{k})|^2\gamma_o^2}\right)^{-1/2}\text{Im}\,[\chi] \tag{A7}$$

In Eq. A6 and A7, $\chi = \left[f^*(\boldsymbol{k})\sum_{i=1}^{3} e^{i\boldsymbol{k}.\boldsymbol{r}_A^l}\frac{\boldsymbol{r}_A^l}{|\boldsymbol{r}_A^l|}\right]$, and $f(\boldsymbol{k}) = \sum_i^3 e^{i\boldsymbol{k}\cdot\boldsymbol{r}_A^l}$ describes the contribution arising form the nearest neighbor atoms shown in Fig. 13. The $x$ and $y$ component of the term $\chi$ gives the $X$ and $Y$ part of the optical matrix element. Parts (b) and (c) of Fig. 4 show contour plots of the optical matrix elements involving transition between direct bands ($v_\alpha \rightarrow c_\beta, (\alpha = \beta)$) and cross bands ($v_\alpha \rightarrow c_\beta, (\alpha \neq \beta)$), respectively. $\alpha$ and $\beta$ represent the band indices and can take values either 1 or 2 as shown by arrows in the band structure of AB-BLG (Refer Fig. 15 (b)). In the first case, the matrix elements vanish at the Dirac points Fig. 4 (b) which is in contradictory to SLG whereas, in the second case, represented in Fig. 4 (c), matrix elements show a maximum exactly at the Dirac points.

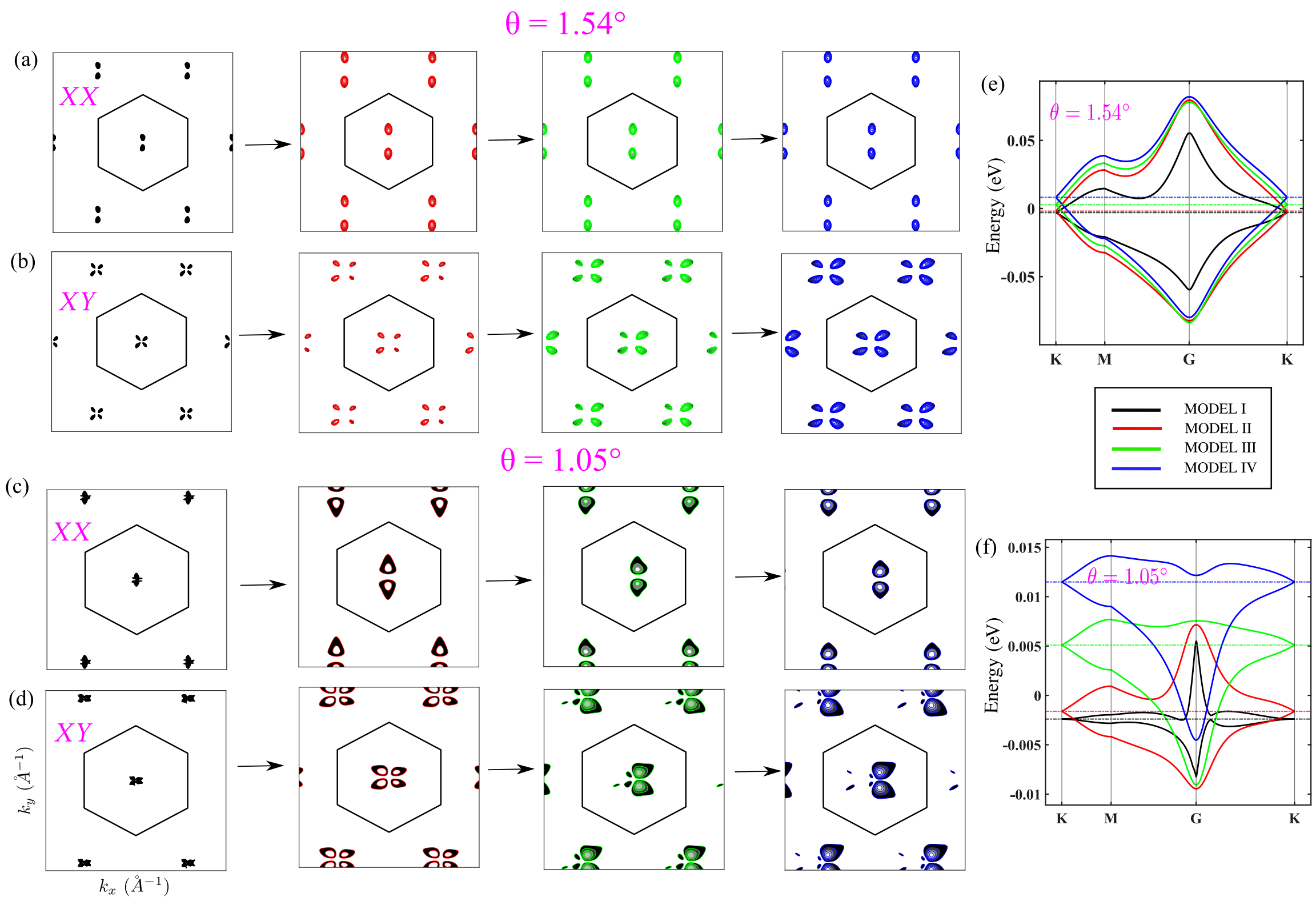

FIG. 17: Contour plots of optical matrix elements $P_{cv}$ as a function of $k_x$ and $k_y$ in the entire MBZ for (a-b) $\theta = 1.54°$ and (c-d) $\theta = 1.05°$. The lowest two bands for these angles are shown in (e) and (f), respectively. Different colors represent different models as mentioned in the legend. Black hexagon outlines the MBZ.

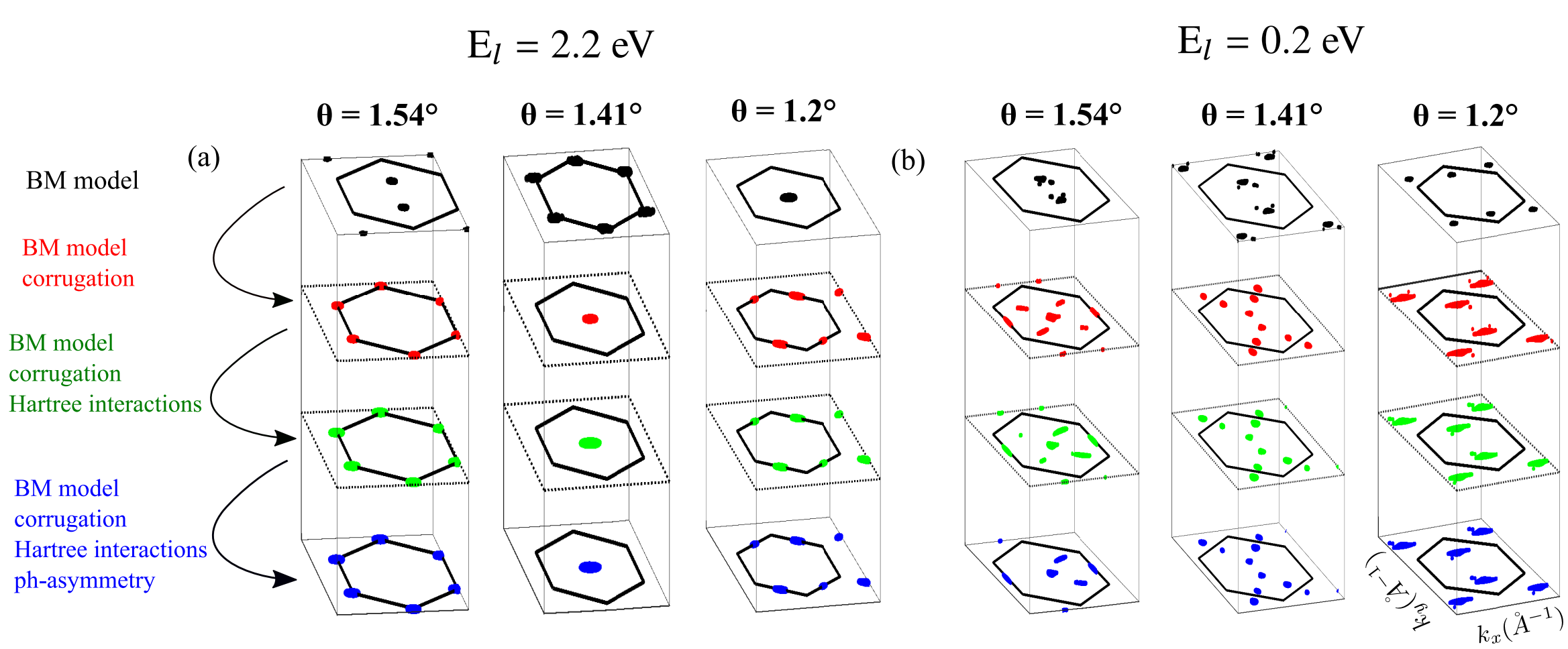

FIG. 18: Transformation of the dominant wave vectors responsible for Rayleigh scattering process ( parallel polarization case : $XX$) on including various effects to the BM model for different twist angles at laser energies $E_l$ (a) 2.2 eV (b) 0.2 eV, respectively. Black hexagon outlines the MBZ.

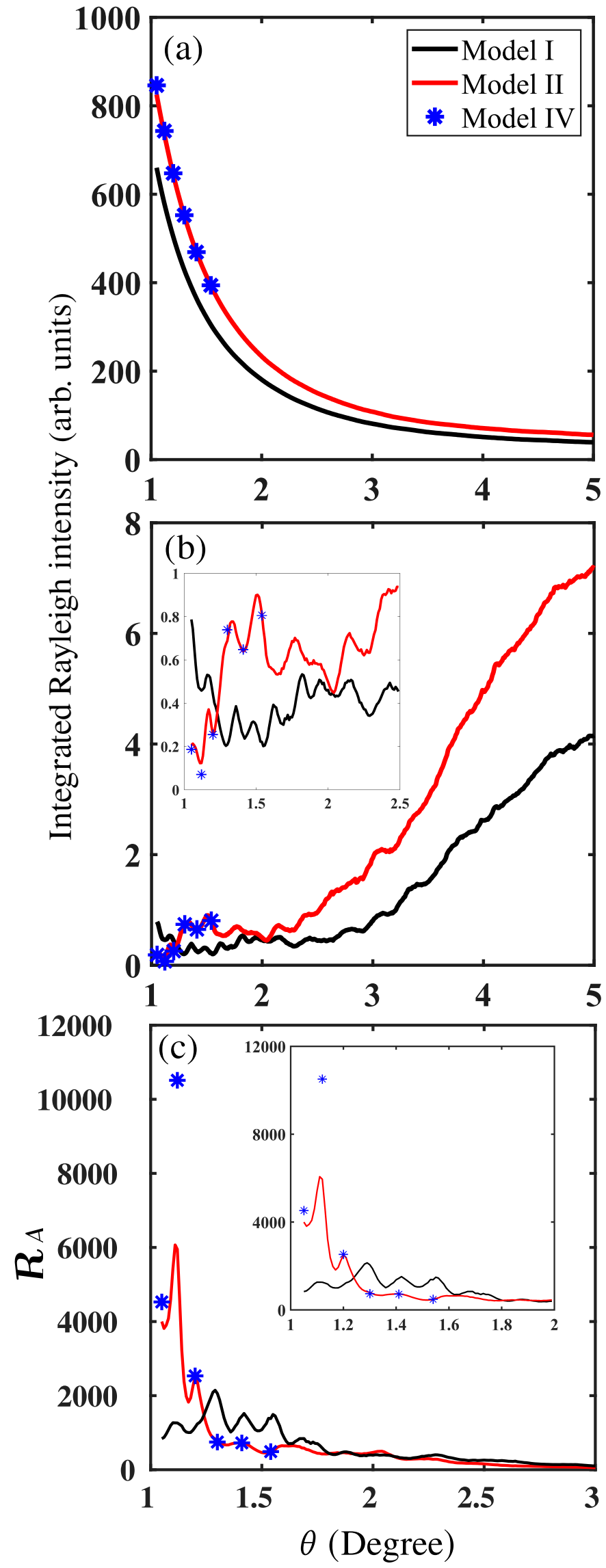

FIG. 19: Integrated Rayleigh intensity (in arb. units) for (a) parallel (b) cross-polarization, respectively. (c) variation of $\boldsymbol{R}_A$ with $\theta$ at $E_l = 2$ eV.

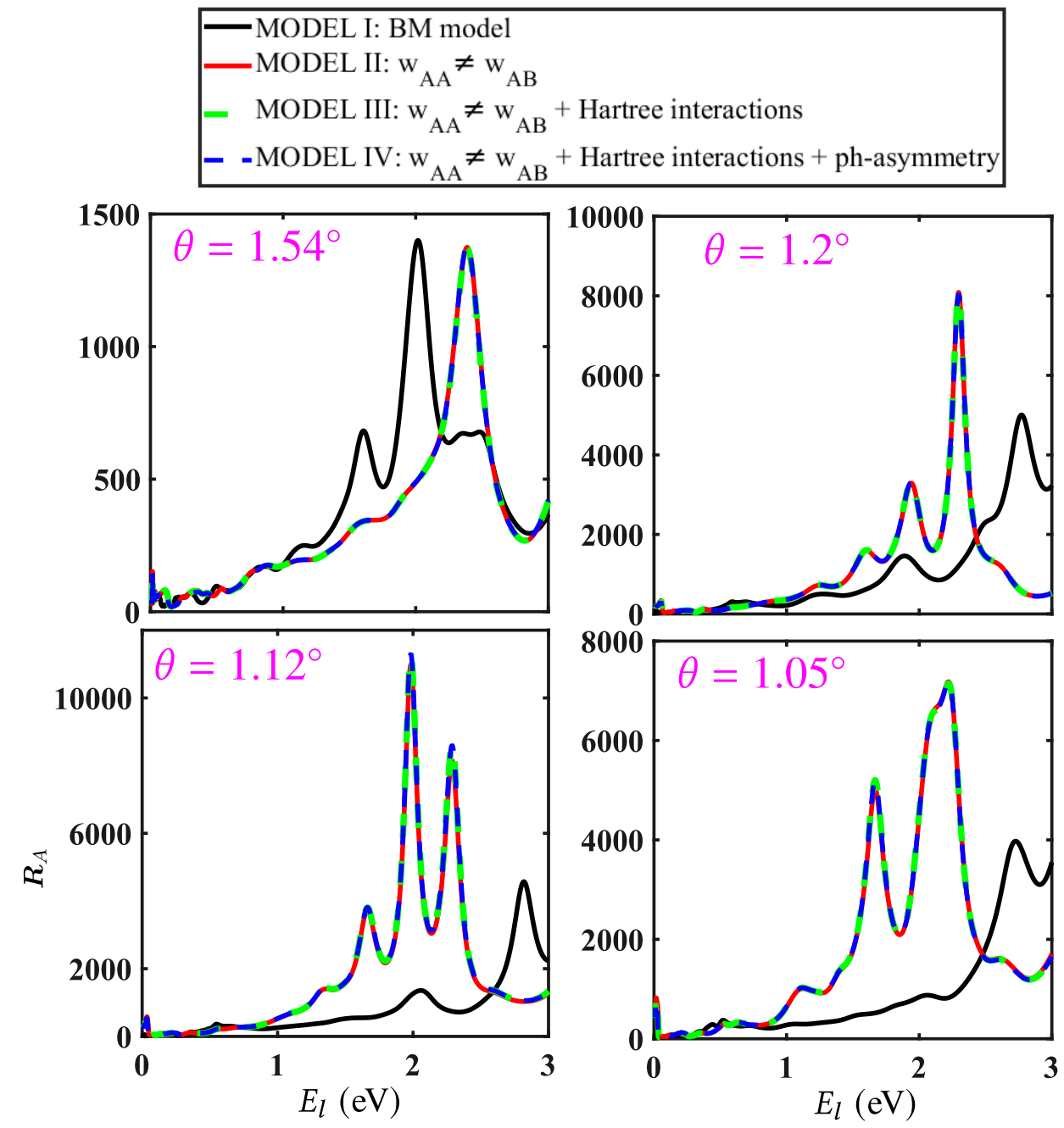

FIG. 20: Variation of the ratio $\boldsymbol{R}_A$ with incoming laser energy $E_l$ for different twist angles.

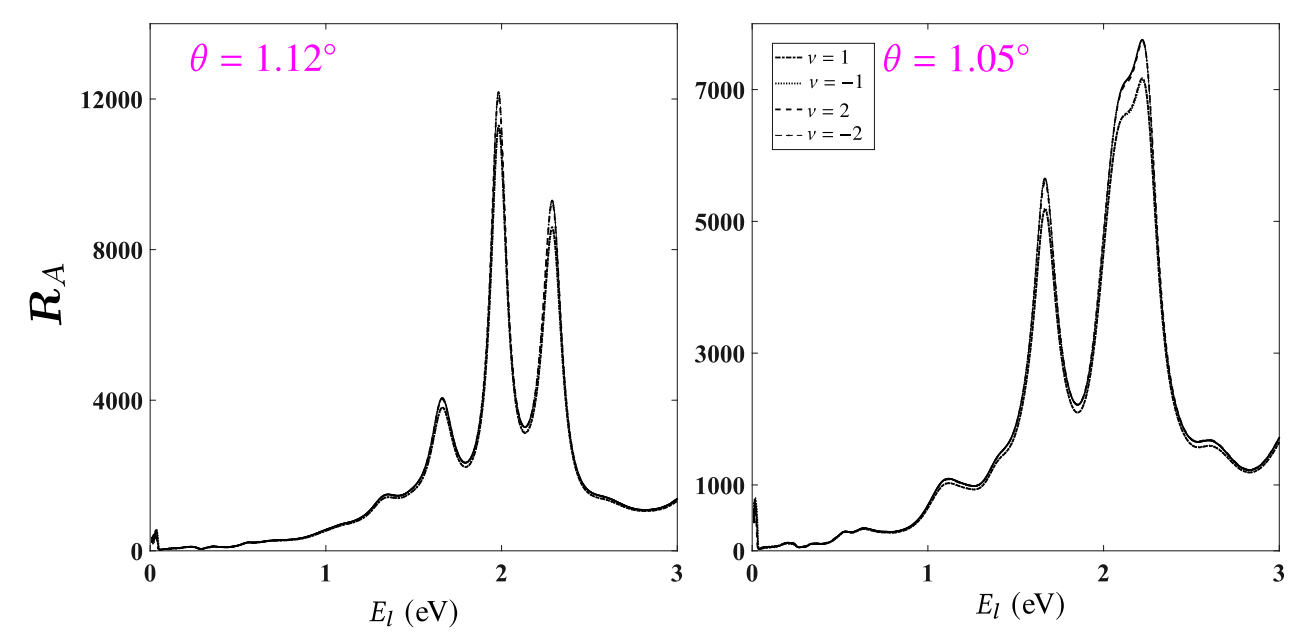

FIG. 21: Effect of doping on ratio $\boldsymbol{R}_A$ as the incoming laser energy $E_l$ is varied for (a) $\theta = 1.12°$, and $\theta = 1.05°$, respectively.

### Appendix B: Step-wise modification to the pristine BM model

We investigate the Rayleigh response of twisted bilayer graphene by making step-wise modifications to the pristine BM model as shown in Fig. 16.

Under light matter interaction, the low energy Hamiltonian described by the BM continuum model [29],

$$H_{BM}(\mathbf{k}) = \begin{pmatrix} h_{\boldsymbol{k}'}\left(\frac{\theta}{2}\right) & T_{\rm b} & T_{\rm tr} & T_{\rm tr} \\ T_{\rm b}^{\dagger} & h_{\boldsymbol{k}'_{\rm b}}\left(-\frac{\theta}{2}\right) & 0 & 0 \\ T_{\rm tr}^{\dagger} & 0 & h_{\boldsymbol{k}'_{\rm tr}}\left(-\frac{\theta}{2}\right) & 0 \\ T_{\rm tl}^{\dagger} & 0 & 0 & h_{\boldsymbol{k}'_{\rm tl}}\left(-\frac{\theta}{2}\right) \end{pmatrix} \quad \text{(B1)}$$

where, $T_j = w_o\sigma_o + w_1\left[\sigma_x \cos\left(\phi'\right) + \sigma_y \sin\left(\phi'\right)\right]$, with $w_o = w_1$. Corrugation effects are included by setting $w_o \neq w_1$ in Eq. B1. Corresponding to the Model III of Fig. 16, Hartree interactions are included by the parametrization provided by Ref. [85] (refer Sec. III for details). The particle-hole transformation operator, $\mathcal{P}$ defined as $i\tau_y$—the interchange of the two layers and changing the sign of the top layer—followed by the in-plane inversion $\boldsymbol{r} \to -\boldsymbol{r}$. Retaining only the contact interlayer terms in the BM model, Kang *et al.* [92] showed that $\mathcal{P}^{\dagger}H_{BM}(\mathbf{k})\mathcal{P} = -H_{BM}(-\mathbf{k})$. This symmetry can be broken by the higher order gradient terms and by introducing the interlayer contact coupling $w_3$ as done in Eq. 7 in the main text.

In Fig. 17, we plot the contour plot of optical matrix elements in the entire MBZ for four models described in Fig. 16 for (a-b) $\theta = 1.54°$, and (c-d) $\theta = 1.05°$, respectively. For these two twist angles, the variation of bands closest to the Fermi level under four different models are shown in parts (e) and (f) of Fig. 17. When the pristine BM model is taken into consideration, the dominant contribution to the Rayleigh scattering process at $\theta = 1.05^\circ$ (magic angle) is due to the wavevectors exactly at the $\Gamma$ point of the MBZ. As soon as corrugation effects are included, the dominant wavevectors move away from the $\Gamma$ point. The location of the dominant wavevectors in the contour plots of optical matrix elements remains essentially unaltered when the Hartree interactions and ph asymmetry is introduced.

The location of the wavevectors responsible for the Rayleigh scattering process for small twist angles is shown in Fig. 18 at $E_l$ = (a) 2.2 eV and (b) 0.2 eV, respectively. We observe that the location of these wavevectors are significantly affected by the corrugation. *e.g.*, for $\theta = 1.41^\circ$, the dominant wavevectors stem from the $\boldsymbol{K}$ point of the MBZ. As soon as corrugation is added to the system, the location of these wavevectors shift to the $\Gamma$ point of the MBZ. The other two additions (Hartree interactions and ph asymmetry) to the system show a negligible effect to the location of the wavevectors, irrespective of the incoming laser energy.

The integrated Rayleigh intensity at $E_l$ = 2 eV is plotted as a function of twist angle $\theta$ in Fig. 19 for (a) parallel, and (b) cross-polarization, respectively, for four different models. The variation of ratio $\boldsymbol{R}_A$ with $\theta$ is plotted in part (c) of Fig. 19. Due to the unavailability of parameters appearing in Eq. 5 at all the twist angles considered, we only provide the available data corresponding to model IV (marked in blue asterisks in Fig. 19). We observe that the blue asterisks show a reasonably good agreement with the data obtained by employing model II.

Fig. 20 plots the variation of ratio $\boldsymbol{R}_A$ with the incoming laser energy, $E_l$, for different models labelled in Fig. 16. It is observed that this ratio is affected only by the corrugation, while other interactions have a negligible effect. The effect of the doping level in the Hartree interaction on the ratio $\boldsymbol{R}_A$ is shown in Fig. 21. It is observed that the position of the peaks remain essentially unchanged on varying the doping level in tBLG.